\documentclass{aa}  
\usepackage{comment}
\usepackage{siunitx}
\usepackage{subcaption}
\usepackage{nicefrac}
\usepackage{graphicx}
\usepackage{color}
\usepackage{hyperref}
\usepackage{csvsimple}

\usepackage{booktabs,wrapfig}
\hypersetup{colorlinks,linkcolor={blue},citecolor={blue}}  

\usepackage{textcomp,gensymb}
\usepackage{caption} 
\usepackage{tabularx}
\usepackage[bottom]{footmisc}
\usepackage[varg]{txfonts}
\usepackage{ulem}
\begin{document} 

    \title{Binary properties of the globular cluster 47\,Tuc (NGC 104)}
    \subtitle{A dearth of short-period binaries}

   \author{Johanna M\"uller-Horn \inst{1,2} \thanks{email: mueller-horn@mpia.de}\and
          Fabian G\"ottgens \inst{1} \and
          Stefan Dreizler \inst{1} \and
          Sebastian Kamann \inst{3} \and
          Sven Martens \inst{1} \and
          Sara Saracino \inst{3} \and
          Claire S. Ye \inst{4}}

   \institute{Institut für Astrophysik und Geophysik, Georg-August-Universit\"at G\"ottingen, Friedrich-Hund-Platz 1, 37077 G\"ottingen, Germany
         \and
             Max-Planck-Institut f\"ur Astronomie, K\"onigstuhl 17, 69117 Heidelberg, Germany
        \and Astrophysics Research Institute, Liverpool John Moores University, 146 Brownlow Hill, Liverpool L3 5RF, UK
        \and
        Canadian Institute for Theoretical Astrophysics, University of Toronto, 60 St. George Street, Toronto, Ontario M5S 3H8, Canada
             }
   \date{Received 14 May 2024 / Accepted 07 December 2024}

 
  \abstract
   {
   Spectroscopic observations of binary stars in globular clusters are essential to shed light on the poorly constrained period, eccentricity, and mass ratio distributions and to develop an understanding of the formation of peculiar stellar objects. 
   47\,Tuc (NGC 104) is one of the most massive Galactic globular clusters, with a large population of blue stragglers and with many predicted but as-yet elusive stellar-mass black holes. This makes it an exciting candidate for binary searches.
   
   We present a multi-epoch spectroscopic survey of 47 Tuc with the VLT/MUSE integral field spectrograph to determine radial velocity variations for 21,699 stars.
  
   We find a total binary fraction in the cluster of $(2.4\pm1.0)\%$, consistent with previous photometric estimates, and an increased binary fraction among blue straggler stars, approximately three times higher than the cluster average. We find very few binaries with periods below three days, and none with massive dark companions. A comparison with predictions from state-of-the-art models shows that the absence of such short-period binaries and of binaries with massive companions is surprising, highlighting the need to improve our understanding of stellar and dynamical evolution in binary systems.
}


   \maketitle
%

\section{Introduction}


Observations of globular clusters (GCs) indicate low binary fractions, ranging from approximately $2\%$ to 30\% \citep[and references therein]{Milone+2012}. Despite their relatively low numbers, binary stars play a central role in the evolution of GCs and are closely linked to their structure and dynamics. 

As a result of mass segregation, binaries tend to accumulate in the centers of the cluster and act as dynamic energy sources that help stabilize the cluster against core collapse \citep{Goodman+1989, Heggie+2006}. Additionally, binaries impact line-of-sight kinematics and affect the observed velocity dispersion, necessitating the accurate determination of binary fractions to avoid overestimating cluster masses \citep{Aros+2021}. 
Within the dense cores of GCs, high rates of stellar encounters facilitate the formation of exotic stellar objects, and low-mass X-ray binaries, millisecond pulsars, blue straggler stars, sub-sub giants, and cataclysmic variables, for example, are thought to represent the late evolutionary stages of close binaries. Exploring binary fractions among multiple populations in GCs \citep{Kamann+2020} can offer insights into their formation environments and initial binary fractions \citep[see][for a review on multiple populations]{Bastian+Lardo2018}. 
Moreover, binaries allow indirect detection of dark massive companions, as demonstrated by recent findings of dormant stellar-mass black holes (BHs) in NGC\,3201 \citep{Giesers+2019} and a BH candidate in NGC\,1850 \citep{Saracino+2023}. Searching for binary systems of luminous stars with massive remnant companions is crucial for constraining the BH retention fraction and understanding the potential of GCs as gravitational wave factories, especially given the wealth of gravitational wave data that is currently being collected \citep{Abbott+2023} and expected in the coming decades \citep{Amaro-Seoane+2017}.


Despite these diverse astrophysical applications of binaries, there have been few detailed spectroscopic studies of binary populations in GCs. This is due to the high stellar numbers and densities in the cluster cores, which make representative spectroscopic observations challenging. Only recently, with the advent of integral field spectroscopy and in particular the ESO MUSE instrument \citep{Bacon+2010}, has it become feasible to carry out representative spectroscopic surveys of GCs and the binary stars therein \citep{Giesers+2018, Saracino+2023b}. Here, a similar approach is applied to the binaries in 47\,Tuc. 

With a total mass close to $10^6\,M_\odot$ \citep{Baumgardt_Hilker2018} and age estimates between $10$ and $12$\,Gyr \citep{Salaris+2002,Hansen+2013,Brogaard+2017,Ye+2022}, the GC NGC\,104, commonly referred to as 47\,Tucanae or 47\,Tuc, is a particularly old and massive MW GC. It has a comparably high metallicity ([Fe/H] = -0.72) with respect to other Galactic GCs, and measured core and half-light radii of 0.36' and 3.17', respectively, at a distance of approximately 4.5\,kpc \citep[2010 edition]{Harris1996}. Due to its relative proximity, brightness, and high galactic latitude (RA, Decl = $00^h 24^m 05.67^s,  -72^{\circ} 04' 52.6''$, \citet{Harris1996}), 47\,Tuc has been a frequent observation target, including extensive coverage from the Hubble Space Telescope (HST) \citep[e.g.,][]{Sarejdini+2007,Anderson+2008}. With a relaxation time at the half-light radius of about 3.6\,Gyr \citep[][2010 edition]{Harris1996}, which implies approximately three elapsed relaxation times, 47\,Tuc is a not very dynamically old cluster. Therefore, the radial decrease in the binary fraction is expected to be only moderately steep \citep{Milone+2012,Ji&Bregman2015}. 

Numerical predictions for the amount of unseen dark mass in the form of BHs in 47\,Tuc range from zero to several hundred BHs \citep{Weatherford+2020,Ye+2022,Dickson+2023}. However, the number of BHs in BH-star binaries expected from simulations of GCs is generally low \citep{Morscher+2015,Chatterjee+2017}. Until now, one candidate for BH in a mass-transfer binary system with a probable carbon-oxygen white dwarf companion has been identified in 47\,Tuc from X-ray and radio data \citep{Miller-Jones+2015,Bahramian+2017}. Alongside the population of BHs, simulations suggest that approximately 1000 neutron stars could be retained in a massive GC like 47\,Tuc \citep{Ye+2022}, though only a small fraction are predicted to be detectable in low-mass X-ray binaries, as millisecond pulsars, or via a tidally captured main-sequence (MS) companion. 

Observations of binaries in 47\,Tuc so far include searches for eclipsing binaries \citep[e.g.,][]{Albrow+2001}, photometric studies of the binary MS \citep{Milone+2012,Ji&Bregman2015}, and the identification of compact binaries by X-ray and radio surveys \citep{Heinke+2005, RiveraSandoval+2018, Paduano+2024}. These approaches have been limited to short-period binaries ($P \lesssim 10\,$d) with high inclinations, MS binaries with high mass ratios ($q > 0.5$), and accreting compact binaries, respectively. The comprehensive spectroscopic analysis presented in this study is sensitive to a wide range of periods ($10^{-1} \lesssim P \lesssim 10^{3}$\,d), mass ratios ($0.1 \lesssim q \lesssim 0.8$), and faint companions and therefore can provide valuable complementary information.


In this work, we explore the binary properties of 47\,Tuc by analyzing the radial velocity (RV) variability of individual stars down to two magnitudes below the MS turn-off of the cluster, corresponding to a lower mass limit of approximately $0.5\,\mathrm{M}_\odot$. Thanks to a statistically significant sample of binaries (708) we can investigate the observed binary fraction in the MUSE field of view (FoV) and provide an estimate of the total binary fraction on the basis of predictions from detailed Cluster Monte Carlo simulations \citep{Ye+2022} of the cluster. 

This paper is structured as follows. In Sect.~\ref{sec:observations} we describe the observations, data reduction, and selection of our final sample. We elaborate the approaches used for binary identification by means of RV variability and orbital parameter inference in Sect.~\ref{sec:methods}.  In Sect.~\ref{sec:cmc_simulation} we introduce a Cluster Monte Carlo simulation of 47\,Tuc that we use as a reference for our observations and to perform mock observations of the GC. The observed binary fraction is presented in Sect.~\ref{sec:binary_fraction} together with an analysis of its radial dependence and its dependence on stellar evolutionary stage. We analyze binary demographics, that is, orbital parameter distributions and peculiar objects, in Sect.~\ref{sec:binary_population}. We discuss possible implications and summarize our findings in Sections \ref{sec:discussion} and \ref{sec:conclusion}. 


\section{MUSE observations}
\label{sec:observations}

\begin{figure}
    \centering
    \includegraphics[width=\columnwidth]{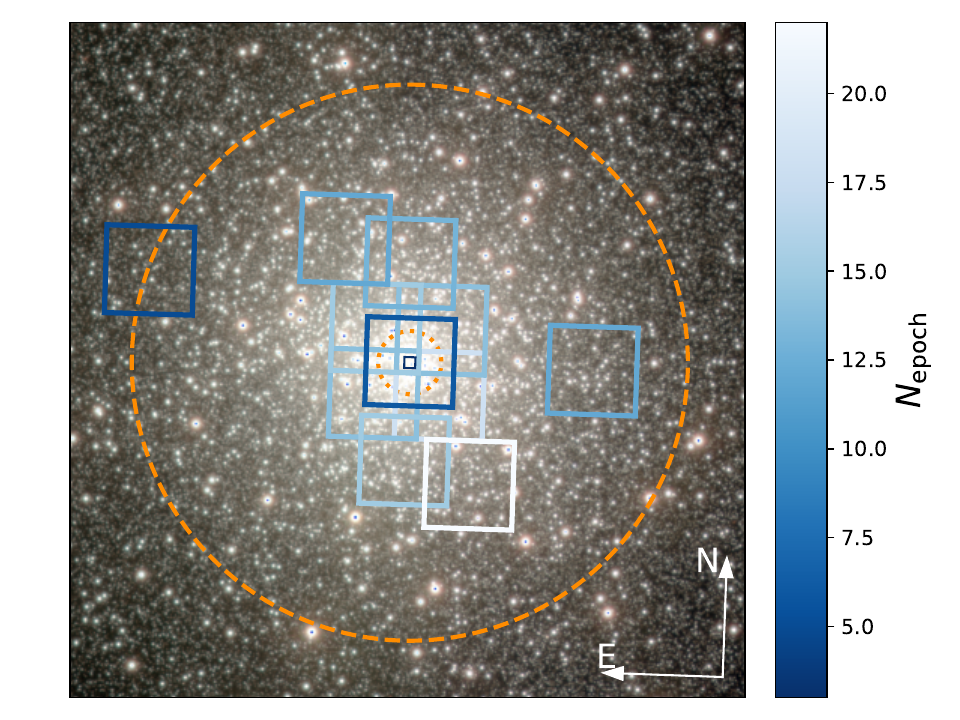}
    \caption{MUSE pointing chart of the globular cluster 47\,Tuc with 11 WFM pointings ($\SI{1}{\arcminute} \times \SI{1}{\arcminute}$) and one NFM pointing color-coded by their respective numbers of observed epochs. The background image is taken from the VMC survey \citep{Cioni+2011}. Half-light and core radii \citep[2010 edition]{Harris1996} are indicated with dashed and dotted lines, respectively.} 
    \label{fig:pointing_chart}
\end{figure}

The observations at hand are part of a larger observational campaign of GCs, the MUSE Galactic globular cluster survey, first presented in \cite{Kamann+2018}. The data were acquired with the MUSE integral field spectrograph, which is mounted on the Very Large Telescope (VLT) at the Paranal Observatory in Chile \citep{Bacon+2010}\footnote{Observing IDs: 094.D-0142(B), 095.D-0629(A), 096.D-0175(A), 097.D-0295(A), 0100.D-0161(A), 0101.D-0268(A), 0102.D-0270(A), 0103.D-0394(A), 0104.D-0257(B), 108.2258.001, 1104.A-0026(E)}. 
MUSE features two operation modes, a wide field mode (WFM) covering a FoV of $\SI{1}{\arcminute} \times \SI{1}{\arcminute}$ with a spatial sampling of \SI{0.2}{\arcsecond} and a narrow field mode (NFM) with a FoV and a spatial sampling of $\SI{7.5}{\arcsecond} \times \SI{7.5}{\arcsecond}$ and \SI{0.025}{\arcsecond}, respectively. Spectra are recorded in the visible wavelength range from 470 to \SI{930}{\nano \meter} with a moderate spectral resolution of $1800 \lesssim R \lesssim 3500$. In WFM observations of GCs, spectra of several thousand stars can be extracted per exposure \citep[see][]{Husser+2016}. 

The observations of 47\,Tuc were carried out between 2014 and 2022 and cover the central region of the cluster to approximately the half-light radius of 47\,Tuc. They consist of 11 partially overlapping pointings in WFM and one central NFM pointing. A pointing chart is shown in Fig.~\ref{fig:pointing_chart}. Each WFM pointing has a minimum of ten visits, with the exception of the central pointing and one outer pointing, which were observed only six times. For each pointing and visit, we obtained three (four for NFM) exposures observed in a dither pattern with derotator offsets of 90 degrees. These are combined in a later step during data reduction to average out possible systematic effects from individual spectrographs. The summed exposure times per pointing ranged from 90 seconds for inner pointings to 2400 seconds (40 minutes) for outer pointings and the NFM pointing. Care was taken to combine only exposures taken within the same 1-hour observing block.

\subsection{Data reduction}
\begin{figure}
    \centering
    \includegraphics[width=\columnwidth]{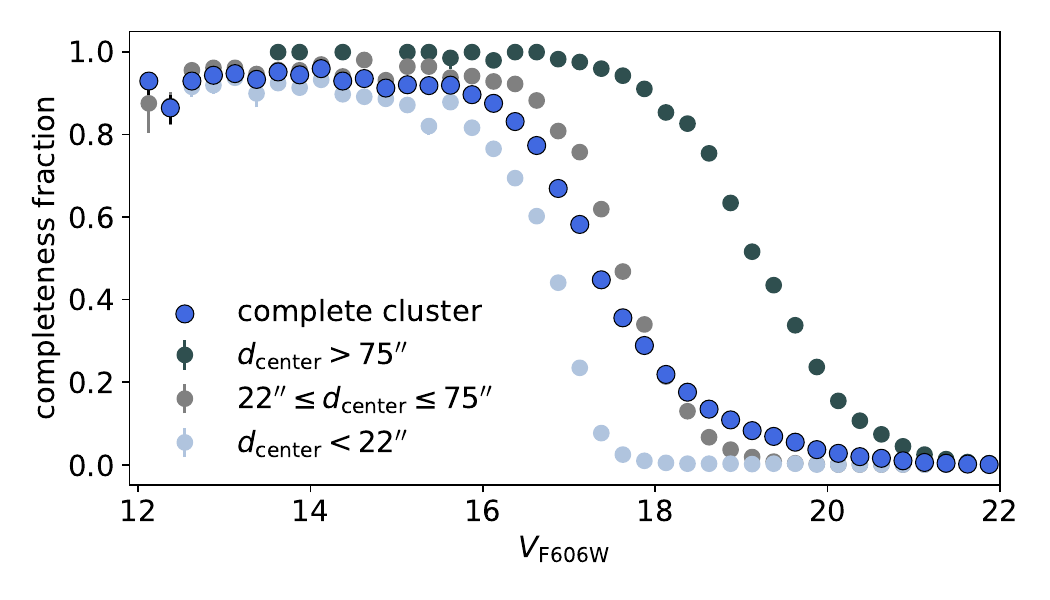}
    \caption{MUSE observational completeness for 47\,Tuc as a function of apparent magnitude and with respect to the HST ACS survey \citep{Sarejdini+2007,Anderson+2008}. The completeness fraction is plotted for pointings in the core, intermediate and outer regions of the cluster as well as the average for the complete cluster. Because of longer exposure times and lower stellar densities, the completeness is higher for larger projected distances from the cluster center, $d_{\mathrm{center}}$.} 
    \label{fig:completeness}
\end{figure}

The  MUSE raw data were reduced using the standard ESO pipeline \citep{Weilbacher+2020}. The extraction of stellar spectra was subsequently performed with \texttt{Pampelmuse} \citep{Kamann+2013}, a fitting software that can disentangle stellar spectra even in crowded regions with non-negligible point-spread-function (PSF) overlap. High-resolution photometric data from the Hubble Space Telescope (HST) ACS survey of globular clusters \citep{Sarejdini+2007,Anderson+2008} were used as reference catalogs of stellar positions and magnitudes. Compared with these catalogs, we computed the observational completeness in the MUSE FoV as a function of apparent magnitude, see Fig.~\ref{fig:completeness}. The average completeness of the observations is high ($>80\%$) for stars of apparent magnitude $\lesssim 16.5$, which roughly corresponds to the MS turn-off point of the cluster, and falls below $10\%$ for apparent magnitude $\gtrsim 19$. Despite the fact that the completeness of observations is lower for MS stars, they still make up the largest part of the data set due to their greater abundance. We note that the completeness limits will vary somewhat between pointings. As a consequence of varying exposure times and stellar densities, the completeness is higher/lower in the outer/central pointings. This is also illustrated in Fig.~\ref{fig:completeness}.    

\subsection{Spectral analysis and RVs}
The spectra extracted from all stars and epochs were fitted to derive stellar parameters (e.g. effective temperature, surface gravity, and metallicity) and RVs of the individual stars. For this purpose, we used the Python package \texttt{spexxy} and performed the fit of the full spectrum based on a grid of synthetic spectra from the \textsc{PHOENIX} library \citep{Husser+2013, Husser+2016}. Initial estimates of the line-of-sight velocity were obtained through cross-correlation with a set of template spectra. The final RV values were determined in a subsequent step employing a Levenberg-Marquardt minimization algorithm. The uncertainties were determined from the variances returned by the spectral fit; the calibration of RV uncertainties is described in more detail in Sect.~\ref{app:radial_velocities} in the Appendix. 
For a comprehensive summary of the methods used in extracting and fitting stellar parameters, see \cite{Husser+2016} and \cite{Kamann+2016,Kamann+2018}. Using telluric absorption features, \cite{Kamann+2016} demonstrats the stability of MUSE's wavelength calibration for night-to-night variations and over the entire FoV, with an accuracy of $\sim 1-2 $\,km/s and shows that reliable RVs can be derived even for spectra with low signal-to-noise ratio (S/N $\gtrsim 5$). 

\subsection{Stellar masses and evolutionary stages}
To analyze binary properties based on evolutionary stage, each star was categorized according to its position in the CMD. The CMD was partitioned into regions representing MS, turn-off (TO), subgiant branch (SGB), red giant branch (RGB), red clump (RC), and blue straggler (BSS) stars. Ambiguous stars near region borders were considered in all applicable evolutionary stages.

Stellar masses were determined using isochrone fits to ACS globular cluster survey photometry \citep{Sarejdini+2007,Anderson+2008}. The best-fit \textsc{PARSEC} isochrone has parameters [M/H] $= -0.7$, age $14.9\,$Gyr, reddening $A_V = 0.03$, and a distance of $4450\,$pc but we find that an isochrone with a more physically motivated lower age yields consistent stellar parameters. Each star in the MUSE data set was assigned an individual mass corresponding to the closest point on the isochrone. BSS were assigned a fixed mass of $(1.20 \pm 0.05)\,M_\odot$, following \cite{Baldwin+2016}.

\subsection{Quality cuts}
\label{sec:sample_selection}
To ensure reliable RVs and uncertainties, a set of quality checks was performed. Only spectra that have S/N $> 5$ were kept in the data set. Spectra which were extracted within five pixels from the edge of a MUSE cube were removed to avoid systematic effects. Spectra for which \texttt{Pampelmuse} failed to deblend multiple sources were also removed, as well as spectra with magnitude accuracy below 90\% \citep[see][for details]{Kamann+2018}. The reliability factor $R$ introduced by \cite{Giesers+2019} was computed for all spectra and the minimum required reliability was set to $R>0.8$.  For each star, nonphysical outliers among RVs, which can occur in cases of low S/N, were identified and excluded following the procedure described in \cite{Giesers+2019}. 

For a blind spectroscopic survey with a FoV spanning several arcmin, it is expected that the sample will be contaminated by foreground stars not belonging to the cluster. The member stars of the GC were identified through expectation maximization \citep{Kamann+2016} based on our measurements of mean RVs and metallicities of the stars. Here, we adopted a Besançon numerical model of the Milky Way \citep{Robin+2013} for the contaminating population. By requiring a membership probability $>0.8$, 1378 likely non-members were removed from the data set. Proper motions from the HST proper motion catalog of Galactic GCs \citep{Libralato+2022} are available for 25,723 stars (approximately 88\% of the cleaned sample). We used these proper motions to independently estimate membership probabilities and thereby assess the reliability of the derived metallicities. For stars with proper motion data, we removed non-members using iterative $\sigma$-clipping with a threshold of $5\sigma$. More than 99\% of the members identified by metallicity and RV cuts were also confirmed as members by this proper motion-based approach. Of the 46 stars not recovered, only two are binary candidates, each with fewer than ten epochs. We conclude that the metallicity-based membership probabilities are reliable and have a negligible impact on the inferred binary properties of the cluster.

Finally, only stars with at least two measured spectra passing all of the above filters were kept in the sample. 

\begin{figure}[t]
    \centering
    \includegraphics[width=\columnwidth]{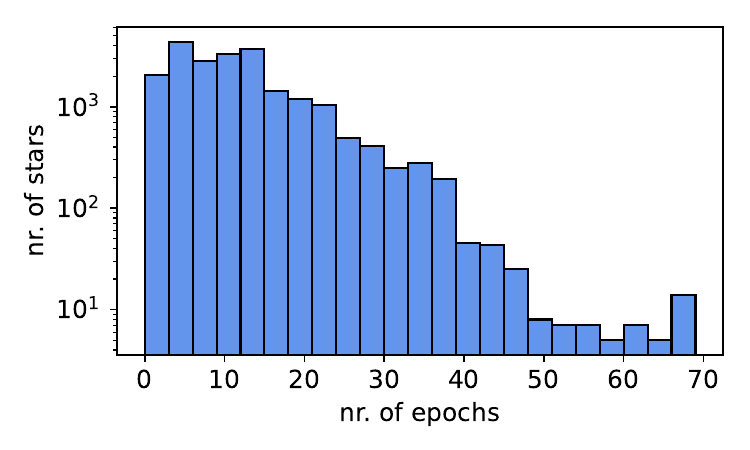}
    \caption{Histogram of the number of observations per star in the filtered MUSE data set comprising 245,522 spectra of 21,699 stars with a median of 11 spectra per star.}
    \label{fig:n_epochs}
\end{figure}


\subsection{Photometric variability}
\label{sec:photometric_variables}
\begin{figure*}[t]
    \centering
    \includegraphics[width=\textwidth]{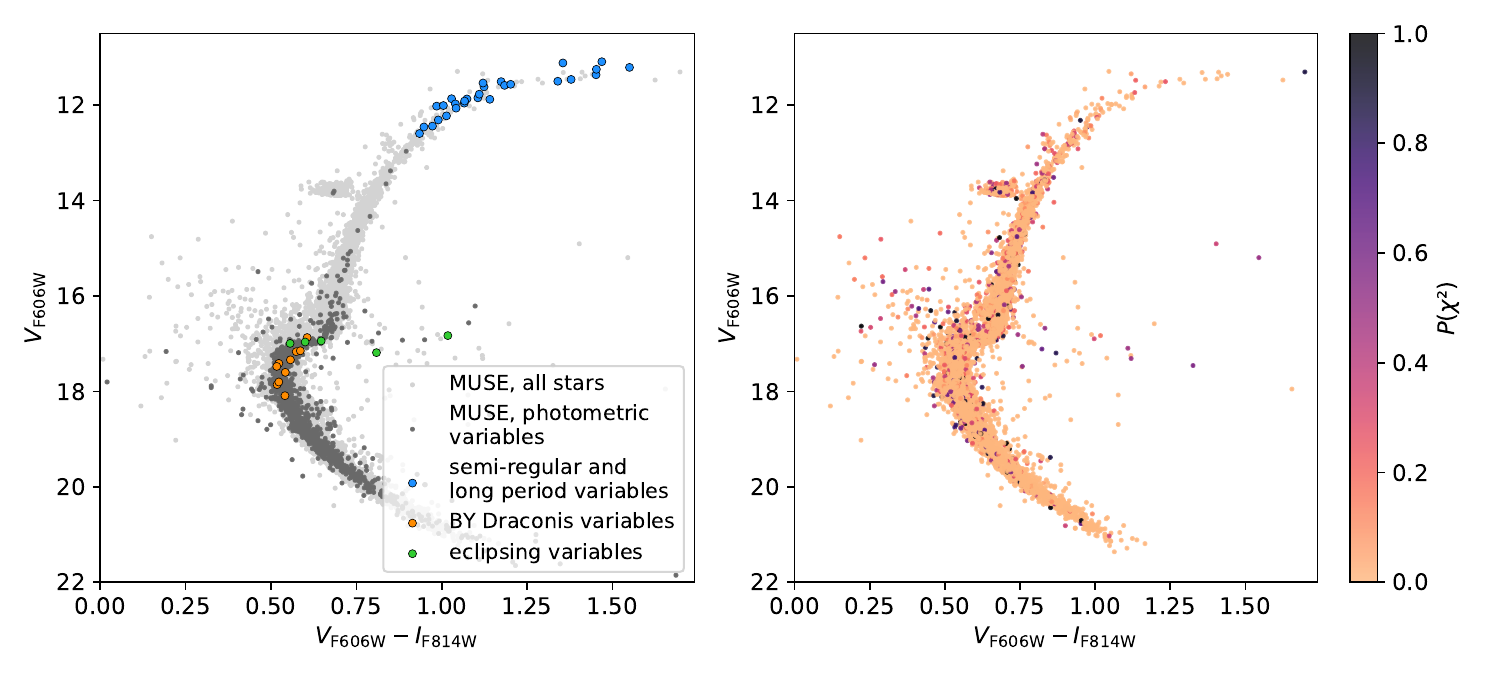}
    \caption{Color-magnitude diagram of the stars contained in the filtered MUSE data set, generated using ACS HST photometry. {Left:} Long-period and semi-regular pulsating variables, BY Draconis-type variables, and eclipsing binaries cross-matched from the catalog of known variable stars in GCs by \cite{Clement+2001} are highlighted in blue, orange and green, respectively. Stars discarded from the data set because of photometric variability in the MUSE data are indicated in dark grey. {Right:} All stars in the MUSE filtered data set color-coded according to their variability probability.}
    \label{fig:photometric_variables}
\end{figure*}

The presence of inherently variable stars that exhibit RV variations, for example due to pulsations and not because of binarity, can in principle lead to an overestimation of the number of binaries. In order to differentiate between true binaries and intrinsically variable stars, the MUSE data set was cross-matched with the catalog of known variable stars in GCs by \cite{Clement+2001}. 39 known photometric variables were excluded from further analysis. These comprise 29 giant branch stars classified as long-period, semiregular, or Mira-Ceti-type variables and ten MS, TO and SGB stars belonging to the group of BY Draconis-type variables. There are five known eclipsing binaries in the \cite{Clement+2001} catalog that could be cross-matched to stars in the MUSE FoV. These remained in the filtered data set. The positions in the CMD of all 44 cross-matched variable stars are highlighted in Fig.~\ref{fig:photometric_variables}, left panel. We also cross-matched our dataset with the Hubble Catalog of Variables \citep{Bonanos+2019}, resulting in 102 matches. These stars were flagged but not removed from the sample, as the source of variability is uncertain and could be caused by binary interactions, such as ellipsoidal modulation, eclipses, or reflection effects.

Stars that exhibit strong photometric variability in the integrated fluxes measured with MUSE were also excluded from the final sample of binary candidates (we required a magnitude accuracy $\geq90\%$, see Sect.~4.4 in \citet{Kamann+2018} for details). These stars are usually not true photometric variables, but spectral variability was accidentally introduced during the extraction process. This may be the case, for example, when bright neighboring stars contaminate the spectrum to varying degrees or when the star is close to the pointing edge and its PSF is only partially covered. The location of these stars in the CMD is also indicated in Fig.~\ref{fig:photometric_variables}, left panel. This filtering is not expected to remove true photometrically variable binaries from the sample, since photometric variability owing to ellipsoidal modulation or partially eclipsing binary stars is below the instrument sensitivity. 

We carefully analyzed the impact of removing photometric variables and in particular the effect of this cut on the inferred binary period distribution. Among the excluded stars only one yielded a well-constrained orbital solution and manual inspection together with a high inferred eccentricity at short orbital period suggests this detection is likely spurious. Additionally, a comparison of RV semi-amplitudes (see Sect.~\ref{sec:discussion}, Fig.~\ref{fig:rv_scatter}) shows minimal differences between the samples with and without photometric variables. In particular, filtering out the photometric variables did not preferentially remove stars with high RV amplitudes from the sample. Thus, we conclude that the exclusion of (potential) photometric variables in this way does not introduce a significant bias in the detection of short-period and or high RV amplitude binaries.

\subsection{The final data set }
The filtered data set comprises 21,699 stars and a total number of 245,522 spectra that span an 8-year baseline. Per star, the median and maximum number of observations are 11 and 68, respectively, with most epochs available for stars in the overlapping central pointings. Figure~\ref{fig:n_epochs} illustrates the quantitative distribution of stars observed for a given number of epochs. The chosen observation strategy prioritizes a nonuniform time sampling of epochs, with time steps between individual observations ranging from hours to years. This approach increases the sensitivity to binaries across a broad range of orbital periods, and conducting multiple observations per night helps to significantly mitigate the risk of temporal aliasing.

\section{Methods for binary identification and orbit determination}
\label{sec:methods}

The individual components of binary stars in GCs usually cannot be visually resolved, even with high-resolution instruments such as HST. Therefore, they appear as single point sources and unless both stars are similarly bright, the extracted MUSE spectra (see Sect.~\ref{sec:observations}) are dominated by the brighter of the two stars. These binary stars are referred to as single-line spectroscopic binaries (SB1), and their orbital motion is reflected in the spectra as time-dependent wavelength shifts. Based on these shifts and, consequently, the changes in inferred RV, it is possible to distinguish between binary stars and single stars.
For a star in a Keplerian orbit, the RV curve as a function of time is given by 
    \begin{equation}
        v_{\mathrm{rad}}(t) = v_{\mathrm{z}} + K \left(\cos(\omega + \nu(t)) + e\cos(\omega)\right)\,. \label{eq:vrad}
    \end{equation}

\noindent Here $\omega$ is the argument of pericenter, $\nu$ is the true anomaly, $e$ is the eccentricity, and $v_z$ is a constant velocity offset. The velocity semi-amplitude $K_1$ of the primary star with mass $m_1$ and induced by the companion star of mass $m_2$ is a function of orbital frequency $f = P^{-1}$, that is the inverse of orbital period, masses, orbital eccentricity and inclination $i$
\begin{align}
        K_1 &=\frac{m_2}{m_1+m_2}  \sqrt[3]{ G(m_1+m_2) 2\pi f} \frac{\sin i}{\sqrt{1-e^2}}\,.  \label{eq:RVamplitude}
    \end{align}
The amplitude increases with larger companion masses and smaller separations of the stars \citep[e.g.,][]{LovisandFischer2010}.

\subsection{Variability probability}
\label{sec:variability_probability}
We identified binary candidates by searching for signs of RV variability and assigned individual binary probabilities to all stars following the method introduced by \cite{Giesers+2019} and explained in more detail in Appendix~\ref{appendix:varprob}. The right panel in Fig.~\ref{fig:photometric_variables} displays the stars in the MUSE filtered data set color-coded according to their variability probability.

The minimum variability probability required to be considered a binary candidate was set to $P(\chi^2) > 0.5$  in this study, as a compromise in terms of completeness and false positives. Thus, the bulk of the single stars are removed while at the same time ensuring that the majority of binaries remain in the sample. There are 708 binary candidates in the MUSE sample that meet this criterion.

\subsection{Reducing effects of contamination}
\label{sec:contamination}
In the course of the analysis, we noticed a need to increase the nominal velocity uncertainties close to bright stars by up to 25\%. This is to account for a potential increase in RV scatter caused by contamination of the spectra by nearby bright stars. We further opted to compute the variability probabilities separately per star and per pointing and then averaging the probabilities per star over all available pointings with at least two measured spectra.\footnote{The NFM pointing and the central pointing were excluded when taking the average since they alone do not span a sufficiently long time to detect variability with periods of weeks to months.} True binary stars should be variable regardless of the pointing in which they were observed. 
We refer to Sect.~\ref{app:contamination} in the Appendix for a more detailed motivation for and description of these measures.

\subsection{Orbital parameter inference}
\label{sec:nested_sampling}

With the aim of inferring binary orbital parameters within a Bayesian framework, we needed to traverse an extensive parameter space where orbital values can span several orders of magnitude. The task is complicated by the oftentimes sparse and noisy data, resulting in complex, multi-modal likelihood landscapes. To explore the posterior probability density functions (PDFs) for the 47\,Tuc binaries, we employed a nested sampling algorithm, known for its ability to robustly sample from complex multimodal distributions \citep{Skilling2004,Skilling2006}. Nested sampling efficiently computes the Bayesian evidence that can be used for model comparison and as a side product produces estimates of the posterior PDFs, making it well-suited for parameter estimation. We worked with the Monte Carlo nested sampling algorithm MLFriends \citep{Buchner2014,Buchner2017} using the \href{https://johannesbuchner.github.io/UltraNest/}{UltraNest} package \citep{Buchner2021}. 

For each binary, we determined the posterior PDFs for six orbital parameters $(K_1, f, M_0, e, \omega, v_z)$, see equations \ref{eq:vrad} and \ref{eq:RVamplitude}, together with an additional jitter term $s$ to account for potentially underestimated uncertainties. $M_0 = 2\pi f t_0$ is the mean anomaly at a reference time $t_0$. The prior distributions adopted for all seven parameters are summarized in Table~\ref{tab:prior_PDF}. 
For the orbital frequency prior the prior range was limited to $10^{-3}\,\mathrm{d}^{-1} \leq f \leq 10^1\,\mathrm{d}^{-1}$ in accordance with the detection limits imposed by time sampling. For the RV semi-amplitude, a normal distribution was adopted that is centered on zero and has a frequency and eccentricity-dependent width $\sigma_K$ given by \citet{Price-Whelan+2020} 

\begin{equation}
    \sigma^2_K(f,e) = \sigma^2_{K,0} \left(\frac{f}{f_0}\right)^{\frac{2}{3}} (1-e^2)^{-1}\,,
\end{equation}
where we set $f_0 = 0.1\,\mathrm{d}^{-1}$ and $\sigma_{K,0} = 30\,$km/s. 
The assumed log-likelihood function has the form 
\begin{equation}
    \log \mathcal{L} = \sum_i -\frac{1}{2} \left(\frac{(v_\mathrm{rad, obs}(t_i)-v_\mathrm{rad, model}(t_i))^2}{\sigma_{v_{\mathrm{rad},i}}^2 + s^2}   - \log{\left(\frac{1}{\sigma_{v_{\mathrm{rad},i}}^2 + s^2}\right)}\right)
\end{equation}
for observed RVs $v_\mathrm{rad, obs}$ with uncertainties $\sigma_{v_{\mathrm{rad},i}}$ and predicted RVs $v_\mathrm{rad, model}$, summed over all available epochs $t_i$ for each star.

\setlength{\extrarowheight}{3pt}
\begin{table}[t]
\caption{Summary of prior PDFs for orbital parameter inference.}\label{tab:prior_PDF}
\centering
\begin{tabularx}{\columnwidth}{c c c c}
\hline
Parameter    & Description & Prior PDF \\
\hline
$K_1$ [km/s] & RV semi-amplitude  & $\mathcal{N}(0,\sigma_K)$ \, [1]\\
$f$ [1/d]     & frequency    &  $\ln f \sim \mathcal{N}(\ln{0.1},2.3)$  \\
$M_0$ [rad] & mean anomaly & $\mathcal{U}(0,2\pi)$ \\
$e$       & eccentricity    & $\beta(0.867,3.03)$ \, [2]\\
$\omega$ [rad] & argument of pericenter& $\mathcal{U}(0,2\pi)$ \\
$s$ [km/s]   & jitter term    & $\ln s \sim \mathcal{N}(-4,2.3)$ \\
$v_z$ [km/s]     & system velocity    & $\mathcal{N}(-18,12.2)$ \, [3]\\
\hline
\end{tabularx}
\tablefoot{$\mathcal{N}(\mu, \sigma)$ indicates a normal distribution with mean $\mu$ and standard deviation $\sigma$, $\mathcal{U}(a, b)$ indicates a uniform distribution between $a$ and $b$, $\beta(a,b)$ indicates a Beta distribution with shape parameters $a$, $b$.
}
\tablebib{
[1] \citet{Price-Whelan+2020}; [2] \citet{Kipping2013}; [3] \citet{Harris1996,Harris2010,Baumgardt_Hilker2018}.}
\end{table}
\setlength{\extrarowheight}{0pt}

In many cases, the posterior PDFs of binary orbital parameters are not well constrained. In particular, the frequency distribution oftentimes does not converge to a single solution but remains highly multi-modal. This can be attributed to aliasing and uncertainties introduced by the sparse and noisy data. For the MUSE observations, we selected the "golden" subset of binaries with well-constrained orbits using the clustering approach also used by \cite{Giesers+2019}. Solutions were considered unimodal if the standard deviation of the periods of the posterior samples satisfies the criterion $\sigma(\ln P) < 0.5$. Bimodal solutions were identified by K-Means clustering with $k=2$ clusters using the \texttt{scikit-learn} package \citep{Pedregosa+2012} and requiring that $\sigma(\ln P) < 0.5$ holds for both clusters separately. In the case of two clusters with one cluster well constrained and containing $>95\%$ of posterior samples, the solution was also classified as unimodal. Posterior distributions with three or more modes were deemed unconstrained, although we note that even for these systems, it is generally possible to exclude large regions of parameter space.

\begin{figure}[h]
    \centering
    \includegraphics[width=\columnwidth]{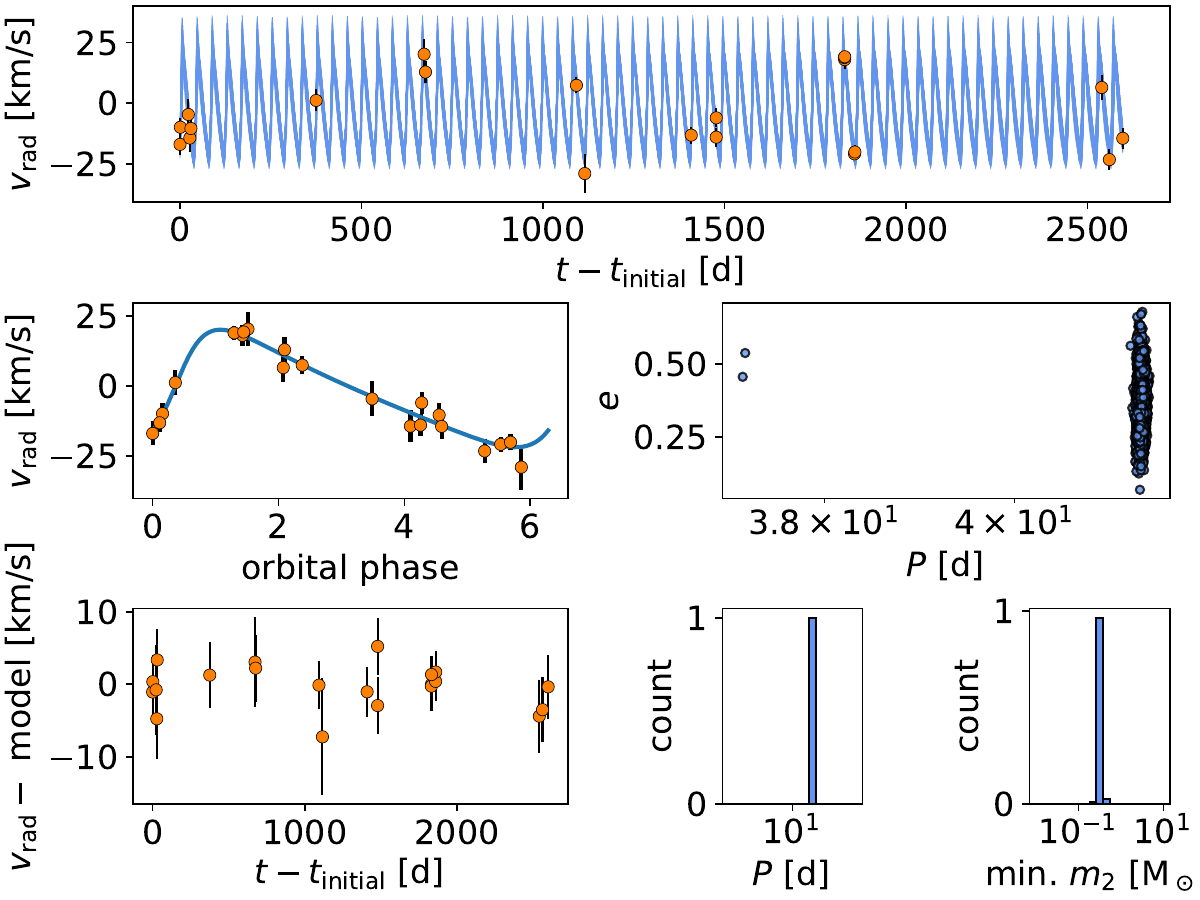}
    \caption{Binary system with a well-constrained unimodal orbit. The upper panel shows the RV time series data (orange markers) and RV orbits computed from the posterior samples (blue lines). The MAP model and phase-folded RVs as well as the residuals from the model are shown in the center and lower left panels. Right panels show the distributions of the posterior samples in period $P$, eccentricity $e$ and minimum companion mass min. $m_2$.}
    \label{fig:MUSE_fit_example}
\end{figure}

In Fig.~\ref{fig:MUSE_fit_example}, the results of one well-constrained binary are displayed, showing the posterior samples of period and eccentricity, the minimum companion mass, and the maximum a posteriori (MAP) phase folded orbit model. The posterior samples are clearly unimodal and clustered around a single orbital solution with a period of 41.4\,d. 
\section{Cluster Monte Carlo simulations of 47\,Tuc}
\label{sec:cmc_simulation}
In this study, we used a simulated population of binary stars from numerical Cluster Monte Carlo (CMC) models of 47\,Tuc as a reference for the observed binary population. The motivation is twofold: first, we can use the simulations to generate mock binary observations (as realistic as possible) that help to evaluate the accuracy and efficiency of binary star detection algorithms. On the basis of these tests, the underlying binary star fraction of the cluster can be inferred from the observed binary fraction. Secondly, the comparison of observations with predictions from simulations is an important step towards a better understanding of these stellar systems. Any discrepancies between models and observations can help inform and constrain the initial conditions for future simulations. 

\subsection{The CMC models}
\label{sec:CMC_models}

\cite{Ye+2022} published a series of CMC models specifically tailored to MW GC 47\,Tuc. These models reproduce cluster properties such as the total mass, core and half-light radii and fit the observed surface brightness and velocity dispersion profiles, pulsar accelerations, and observed numbers of compact objects of 47\,Tuc. They represent the current best matching CMC model for 47\,Tuc, which is why we chose them as a reference for our observations. For a detailed description of the CMC code, we refer to \citep{Rodriguez+2022}. 
For the CMC simulation of 47\,Tuc, the metallicity was fixed to $Z = 0.0038$ according to the value provided by the catalog \cite{Harris1996}, 2010 edition. The initial number of stars was set to $3 \times 10^6$ and an initial binary fraction of $f_b = 0.022$ was implemented following \cite{GierszandHeggie2011}. 
The binary companion masses were initially drawn from a flat distribution in the mass ratio $q$, in the range 0.1–1 based on \cite{Duquennoy+1991}. Binary orbital periods were drawn from a flat distribution in the logarithmic scale from Roche Lobe overflow to the hard/soft boundary. For binary eccentricities, a thermal distribution was used \citep{Heggie1975}. A more detailed description of the simulation setup can be found in \citep{Ye+2022}.

\subsection{Mock RV observations}
\label{sec:CMC_mock_data}

To generate realistic mock data, we employed time patterns of stars observed in 47\,Tuc. Each simulated star was paired with a reference star from the MUSE database, matched in mass and stellar type. Using the binary parameters from CMC, namely the semi-major axis, masses, and eccentricity, we computed RVs at the observation times of the reference star using Eq.~\ref{eq:RVamplitude}. Orbital angles were randomly drawn from uniform distributions, and the inclination was sampled from a uniform distribution in $\cos{(i)}$. An offset in velocity was added to the RVs based on the central velocity dispersion of the cluster. Additional noise was added to the synthetic data within the uncertainties taken from the corresponding MUSE measurements of the reference star. The RV amplitudes were multiplied with an attenuation factor 
$\left(1 - \nicefrac{F_2}{F_1}\right)\,,$ as was done by \cite{Giesers+2019}. This is to account for the fact that the RV amplitude measured with MUSE is linearly damped by the flux ratio $\frac{F_2}{F_1}$ of the stars. For twin binaries with similarly bright stars, the spectral lines cannot be individually resolved and zero RV amplitude is measured. The fluxes were approximated with the help of mass-luminosity and mass-radius relations taken from \cite{Eker+2018}. 

Of all 35368 simulated binaries in the best-fit CMC simulation snapshot at $10.416\,$Gyr, we selected the subset of hypothetically observable binaries according to our observational completeness function for 47\,Tuc (see Fig.~\ref{fig:completeness}) and excluded binaries outside of the MUSE FoV. Within the reduced sample of 1555 simulated binary stars, some exhibit minimal to no RV variability. This may occur for wide orbits, small companion masses, low inclinations, or if the mock data were generated at similar orbital phases. Similarly to the MUSE data binary selection, the final hypothetically observable set of simulated binaries was determined by requiring a variability probability of $P(\chi^2) > 0.5$, leaving 839 simulated binaries.

\subsection{Properties of the synthetic binary population}
\label{sec:CMC_binary_population}

\begin{figure*}[ht!] 
  \begin{minipage}[b]{0.245\textwidth}
    \centering
    \includegraphics[width=\textwidth]{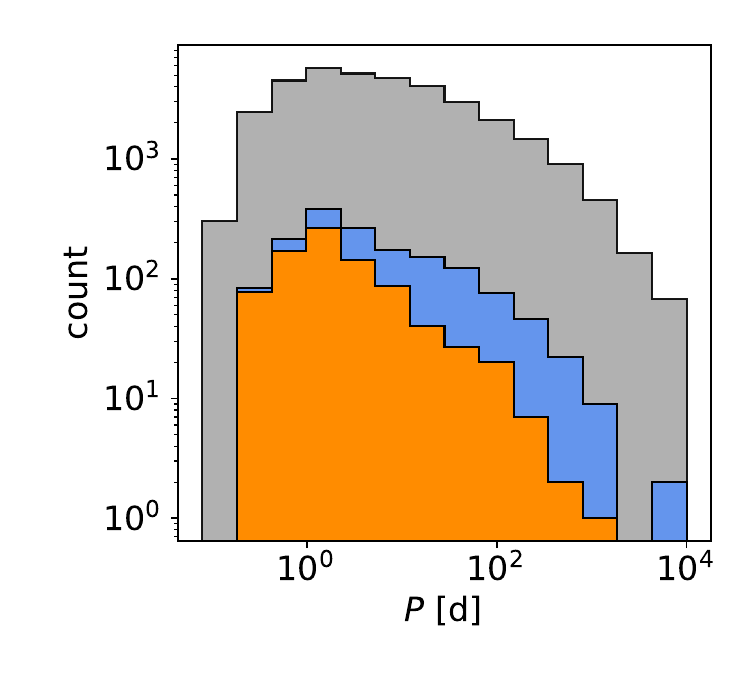} 
  \end{minipage}
  \begin{minipage}[b]{0.245\textwidth}
    \centering
    \includegraphics[width=\textwidth]{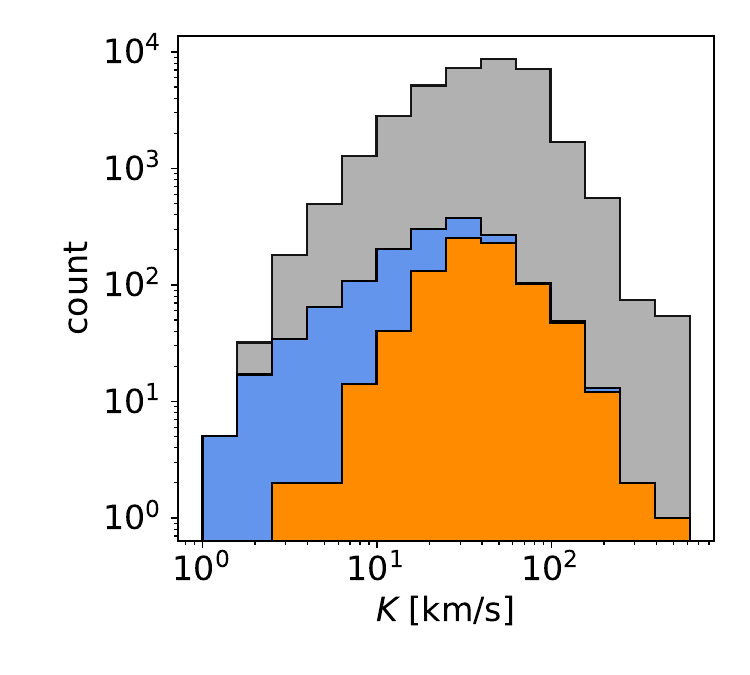} 
  \end{minipage}
  \begin{minipage}[b]{0.245\textwidth}
    \centering
    \includegraphics[width=\textwidth]{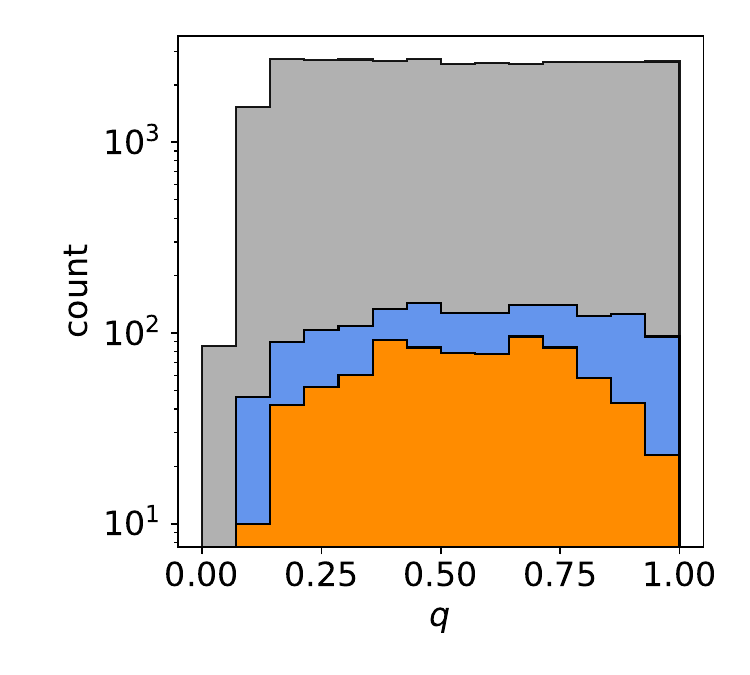} 
  \end{minipage} 
  \begin{minipage}[b]{0.245\textwidth}
    \centering
    \includegraphics[width=\textwidth]{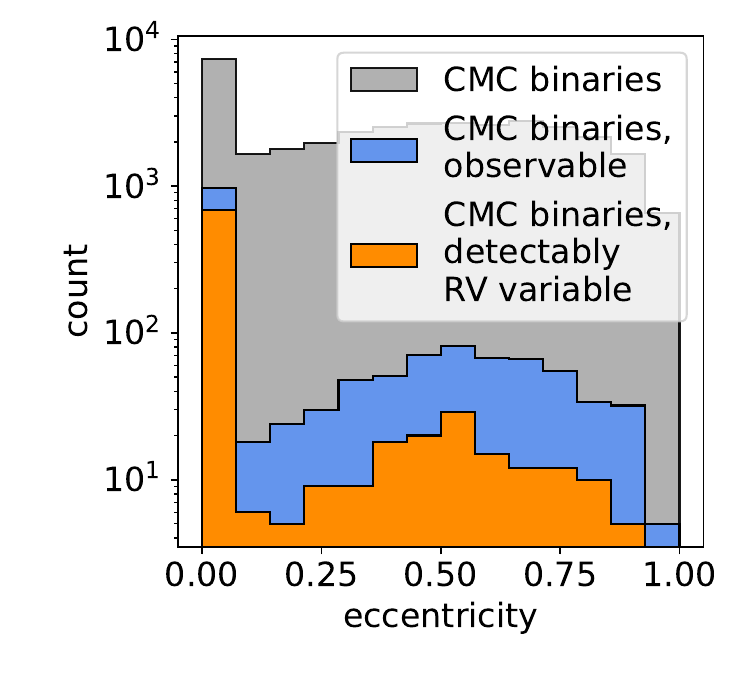} 
  \end{minipage}
  \caption{Histograms of the orbital parameters for the GC 47\,Tuc derived from the CMC simulation. The panels show the orbital period, the RV semi-amplitude, the mass ratio and the eccentricity distributions for all simulated CMC binaries (gray), the hypothetically observable subset in the FoV (blue) and the final subset of detectable RV variable CMC binaries (orange).} 
  \label{fig:mock_orbital_params} 
\end{figure*}

\begin{figure}[t!]
    \centering
        \includegraphics[width=\columnwidth]{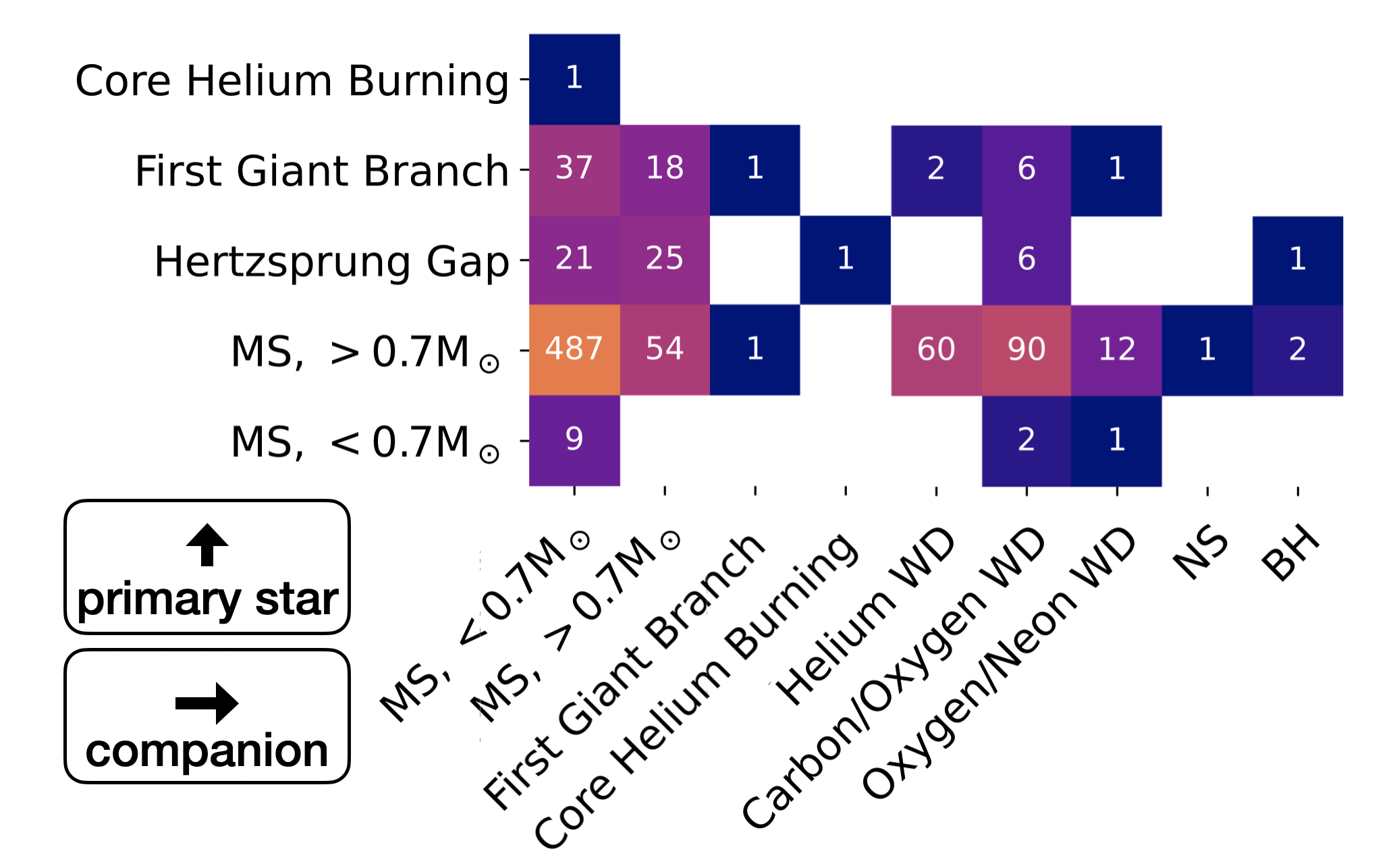}
    \caption{Stellar types of the 839 binary systems in the CMC simulation of 47\,Tuc for the filtered detectable subset. The primary stellar type is plotted on the $y$-axis vs. the stellar type of the companion on the $x$-axis. The white digits indicate the number of binaries with the specified types.}
     \label{fig:ktype_heatmap}
\end{figure}

Histograms of orbital periods, eccentricities, and RV semi-amplitudes of the CMC synthetic binaries are plotted in Fig.~\ref{fig:mock_orbital_params} and visualize our selection function. Of the 35368 binary stars in the CMC simulation (shown in gray), 1555 are hypothetically detectable (shown in blue), meaning that they have a bright enough primary to be observed with MUSE and are located in the FoV. A subset of 839 binary stars is observably RV variable (that is, $P(\chi^2)>0.5$) (displayed in orange), of which 572 in turn have ten or more epochs.
The general shapes of the parameter distributions are similar for the raw, entire set of simulated binary stars and for the subsets of simulated detectable and simulated observably variable binaries. The simulated distribution of the mass ratios is nearly flat, reflecting the initial conditions of the simulations. The synthetic eccentricity distribution has a peak at near-zero eccentricity for circularized orbits and is approximately flat for higher eccentricities. The predicted simulated distribution of orbital periods has a strong peak at short periods of only a few days. The RV semi-amplitudes are approximately log-normally distributed. We note here that the hypothetically observable subset has slightly lower RV semi-amplitudes than the overall raw sample. This is due to the large fraction of simulated binary stars with a low-mass MS primary, which have high RV amplitudes but are too faint to be observed. In Fig.~\ref{fig:ktype_heatmap}, the distributions of stellar types among the 839 hypothetically detectable simulated binaries are shown. The majority of simulated binaries are made up of MS-MS binaries. There is a predicted total number of 184 binaries ($\sim 22\%$) that have a luminous primary and a stellar remnant companion, that is, a white dwarf, neutron star, or BH.

\begin{figure}[t]
    \centering
    \includegraphics[width=\columnwidth]{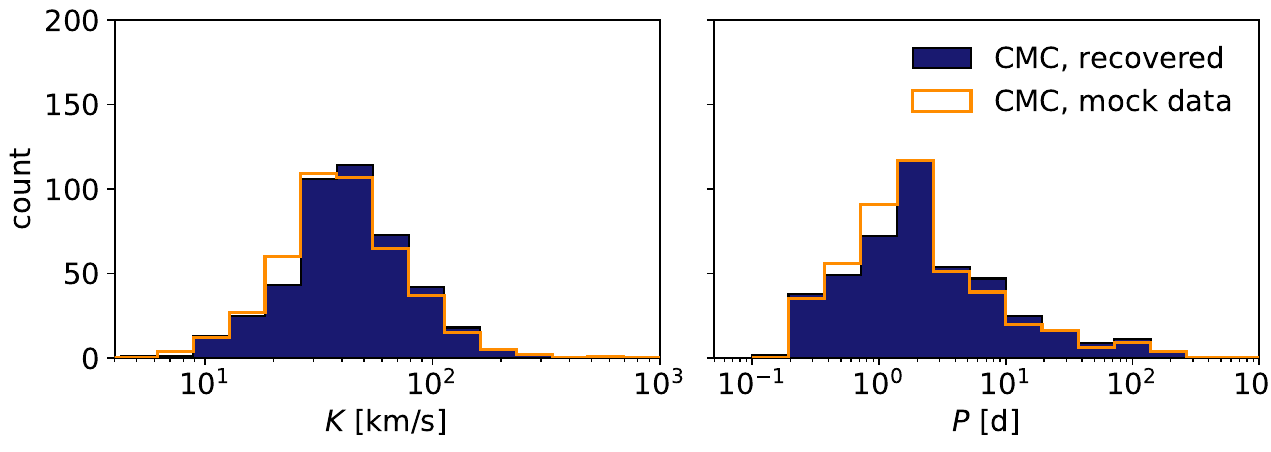}
    \caption{Histograms of inferred RV semi-amplitudes and periods for the subset of 444 simulated detectable binaries in 47\,Tuc with ten or more epochs and well-constrained orbits. The inferred parameter distributions are shown in dark blue, the true input values are indicated in orange.}
    \label{fig:47Tuc_simulated_hist}
\end{figure}

To evaluate the precision of the nested sampling method in estimating orbital parameters, we applied it to the data set comprising 572 detectable CMC binaries with a minimum of ten epochs. The selection of prior PDFs and sampling hyperparameters is detailed in Sect.~\ref{sec:nested_sampling}. Among the 572 simulated binary stars, 444 (78\%) exhibit well-constrained orbital solutions. Their distributions of orbital parameters are also shown in Fig.~\ref{fig:47Tuc_simulated_hist}. The dark blue histograms represent the inferred RV semi-amplitudes and orbital periods, derived from the combined and weighted posterior distributions, while the orange histograms show the true input parameters of the CMC mock data set.

For simulated binaries, RV uncertainties cannot be over- or underestimated, so the jitter term should theoretically be zero. However, we allowed for a jitter term in our parameter inference. This serves as a plausibility check, as the inferred jitter values are strongly peaked at zero, with over 80\% of binaries having a median jitter below 1\,km/s, and over 90\% below 2\,km/s.

The high degree of similarity between the true and predicted histograms of well-constrained binary orbits suggests that the nested sampling method reliably determines the distributions of orbital parameters. 
The subset of well-constrained binaries tends to contain the binaries with higher RV semi-amplitudes and, in turn, higher companion masses. This correlation is expected, as orbits with small RV amplitudes that are comparable to the RV uncertainties exhibit greater degeneracy and are therefore less likely to yield well-constrained fits. Recognizing this slight bias toward more massive, high RV amplitude binaries is crucial for interpreting the companion mass distribution of the observed MUSE binaries.

\subsection{Estimating MUSE sensitivity}
\label{sec:sensitivity}

We investigated the sensitivity of our study and its dependence on the properties of the binaries to better interpret the parameter distributions of the observed binaries, such as the binary fractions among different stellar evolutionary stages (Sect.~\ref{sec:binary_fraction_types}) and the period distribution (Sect.~\ref{sec:period_distribution}).

\begin{figure*}[t]
\centering
\begin{subfigure}{.25\textwidth}
    \centering
    \includegraphics[width=\textwidth]{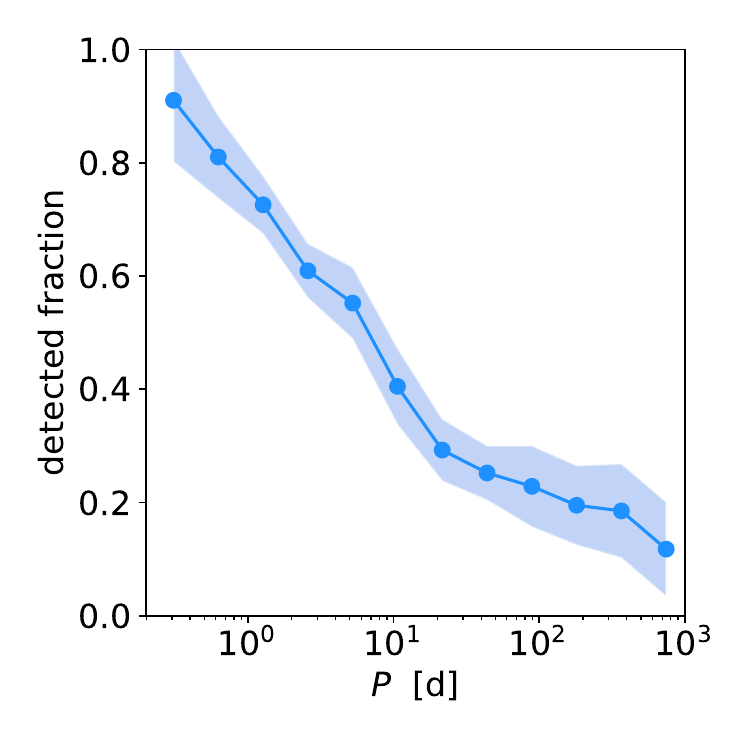} 
\end{subfigure}%
\begin{subfigure}{.25\textwidth}
    \centering
    \includegraphics[width=\textwidth]{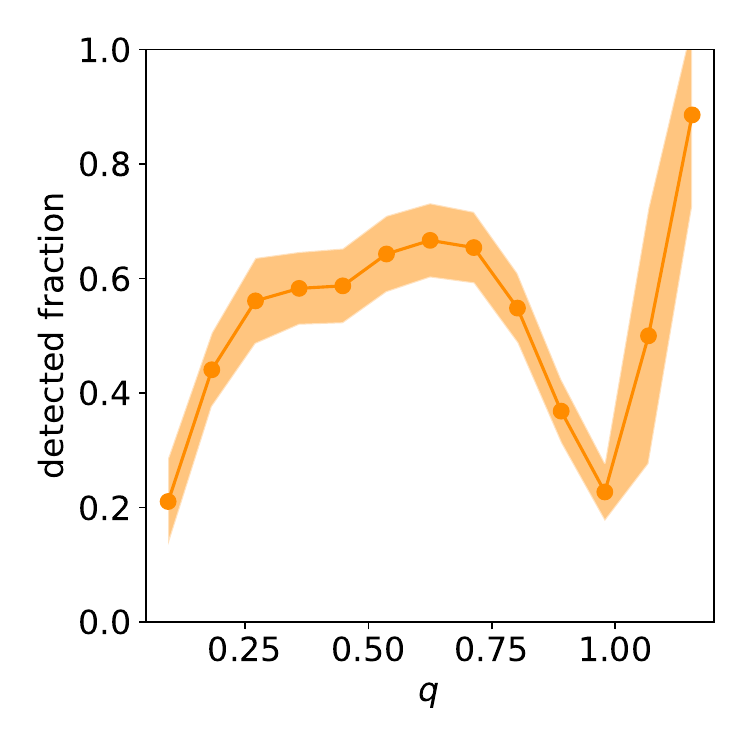} 
\end{subfigure}%
\begin{subfigure}{.25\textwidth}
    \centering
    \includegraphics[width=\textwidth]{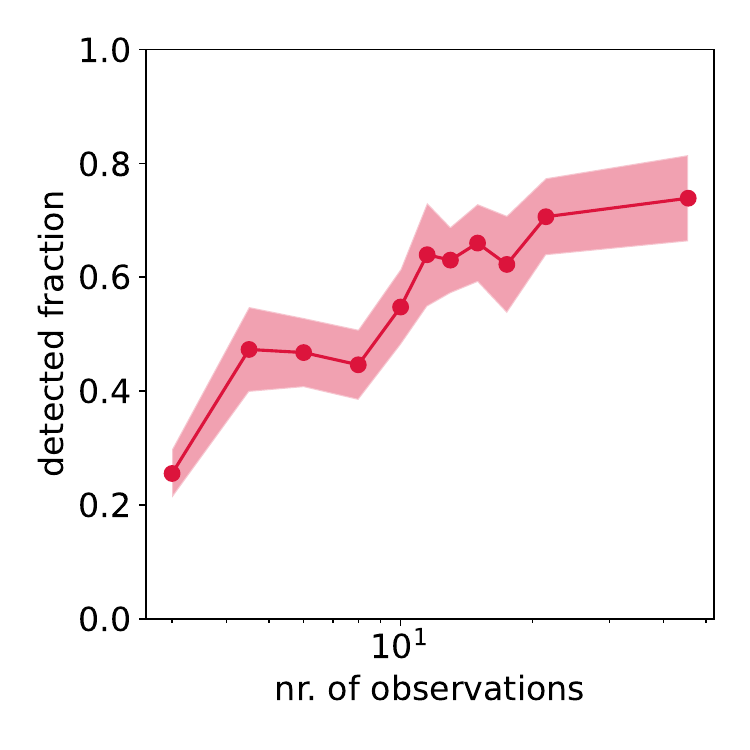}
\end{subfigure}%
\begin{subfigure}{.25\textwidth}
    \centering
    \includegraphics[width=\textwidth]{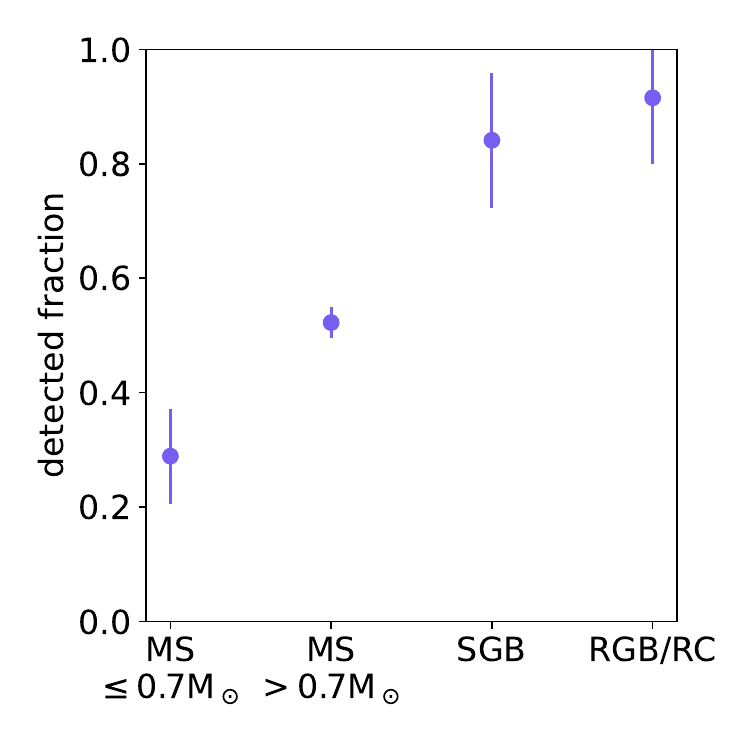}
\end{subfigure}
\caption{Detected fraction of RV-variable simulated binaries with variability probability 
$P(\chi^2) > 0.5$ among all observable CMC binaries, shown as a function of orbital period (left), mass ratio (center left), number of epochs (center right), and stellar evolution stage of the primary (right). Confidence intervals, calculated by varying the probability threshold between 0.4 and 0.6 and adding Poisson noise uncertainties in quadrature, are shown as shaded regions.} 
\label{fig:detection_fraction}
\end{figure*}

The mean detection probability in the FoV, that is, the fraction of true binaries that are identified as such based on their RV variability, is $\frac{839}{1555}\approx 54\%$. 
However, as illustrated in Fig.~\ref{fig:detection_fraction}, the detection probability varies significantly with binary properties. Binaries are easiest to detect if they have short periods implying high RV amplitudes. The detection probability decreases from greater than 80\% for periods up to one day to less than 30\% for periods over one year. Additionally, a higher number of observations slightly boosts the detection probability, though the effect is less pronounced than the dependence on the orbital period. In principle, we are more sensitive to higher mass companions, which induce higher RV amplitudes. The sensitivity is highest for dark massive companions with $q>1$, where $q$ is always the ratio of companion mass to observed primary star mass. However, for unevolved binaries with similarly bright stars (SB2 binaries), observed RVs may be damped or canceled out, leading to a decrease in the detection probability for $q \in [0.7,1.0]$. Finally, there is a notable correlation between the probability of detection and the evolutionary stage of the primary (visible) component of the binary. This is effectively a correlation with S/N as brighter, more massive, and more evolved stars generally exhibit lower RV uncertainties, facilitating the identification of binaries.

\section{Observed binary fraction}
\label{sec:binary_fraction}


\subsection{Binary candidates}
\label{sec:binary_candidates}
\begin{figure}[t!]
    \centering
        \includegraphics[width=\columnwidth]{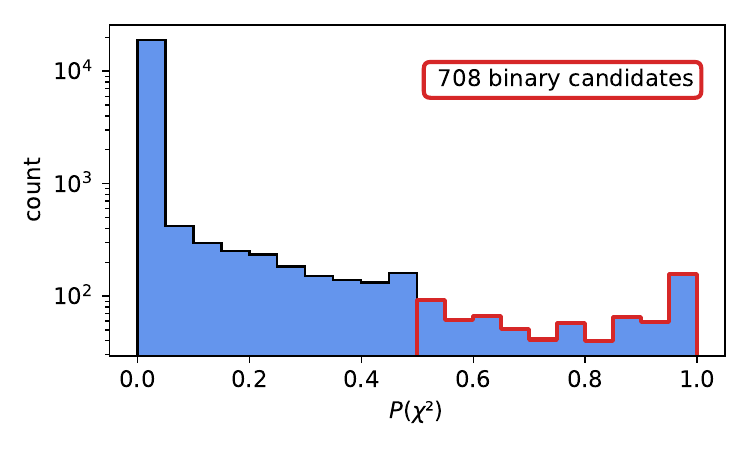}
    \caption{Variability probabilities \citep{Giesers+2019} for all stars in the cleaned RV data set of 47\,Tuc computed per pointing. 
    }
    \label{fig:varprob_hist_and_CDF}
\end{figure}

The refined MUSE data set includes 21,699 reliable member stars with a total of 245,522 spectra and a median of 11 epochs per star. Binary candidates were identified based on individual binary probabilities, as detailed in Sect.~\ref{sec:variability_probability}. Using this statistical approach yields 708 binary candidates, corresponding to a discovery binary fraction of $$f_{\mathrm{bin, discovery}} = (3.3 \pm 1.1)\%$$ in the MUSE FoV. In Fig.~\ref{fig:varprob_hist_and_CDF}, the distribution of binary probabilities is shown, exhibiting a clearly bimodal form. 

\subsection{Binary fraction among different stellar types}
\label{sec:binary_fraction_types}

\begin{figure}[t!]
    \centering
    \includegraphics[width=\columnwidth]{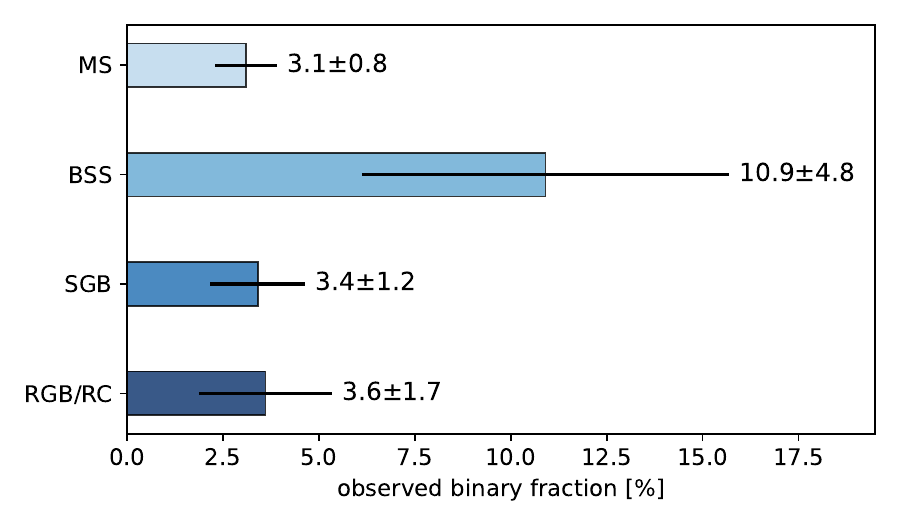}
    \caption{Observed binary fractions across various stellar evolutionary phases, arranged from top to bottom: main-sequence and turn-off stars, blue straggler stars, subgiant branch stars, red giant branch and red clump stars. The uncertainties of the derived fractions, determined by varying the detection threshold and considering the finite sample sizes, are shown as black lines. 
    }
    \label{fig:binary_fraction_stellar_types}
\end{figure}

Fig.~\ref{fig:binary_fraction_stellar_types} shows the observed binary fractions for different stellar evolution stages, including MS, TO, BSS, SGB, RGB, and RC stars. The uncertainties combine statistical Poisson noise and detection uncertainties, which were treated as independent and added in quadrature. Detection uncertainties were estimated by varying the variability probability threshold between 0.4 and 0.6. 
With the exception of the BSS, the binary fractions are consistent across all evolutionary phases, showing no significant variation and remaining low overall. 

Due to the varying completeness of our sample across different evolutionary stages (see Sect.~\ref{sec:sensitivity}), in particular the underlying binary fraction for MS stars may be slightly higher. However, we caution against directly correcting for this by using the inverse discovery probabilities from Fig.~\ref{fig:detection_fraction}, as those reflect the recovered fraction of true binaries. Without accounting for false-positive detections, such a correction would likely lead to a spuriously high inferred binary fraction.

The BSS population shows an elevated binary of $(10.9 \pm 4.8)\%$, roughly three times the cluster average, with 12 out of 110 stars showing RV variability. This elevated binary fraction aligns with the commonly accepted theory that BSS form via mass transfer in compact binaries or multiple systems \citep{McCrea1964}, or through stellar collisions. 

Quantitatively, \cite{Hypki&Giersz2017} predict a binary-to-single star ratio for BSS of $R_{\mathrm{B/S}} \approx 0.4$ from numerical simulations. However, in 47 Tuc, we measure $R_{\mathrm{B/S}} = 0.12$, which is lower than the simulation prediction and much lower than the value of $R_{\mathrm{B/S}} = 1.35$ observed in NGC\,3201 \citep{Giesers+2019}. Some level of incompleteness with respect to the simulations is expected. Recent findings \citep[e.g.,][]{Dattatrey+2023} suggest that many BSS have stripped, hot, low-mass companions. Such companions would induce minimal RV variability in the BSS, making them difficult to detect in our survey. The difference in $R_{\mathrm{B/S}}$ between 47\,Tuc and NGC\,3201 could be due to the overall smaller binary fraction in 47\,Tuc. Alternatively, it might suggest a higher fraction of BSS formed via collision rather than mass transfer in 47\,Tuc compared to NGC\,3201, which is conceivable given the higher stellar densities in 47\,Tuc. Further analysis of the orbital characteristics of individual BSS is provided in Sect.~\ref{sec:peculiar_objects}. 

\subsection{Radial dependence of the binary fraction}
\label{sec:binary_fraction_radial}

\begin{figure}[t]
    \centering
    \includegraphics[width=\columnwidth]{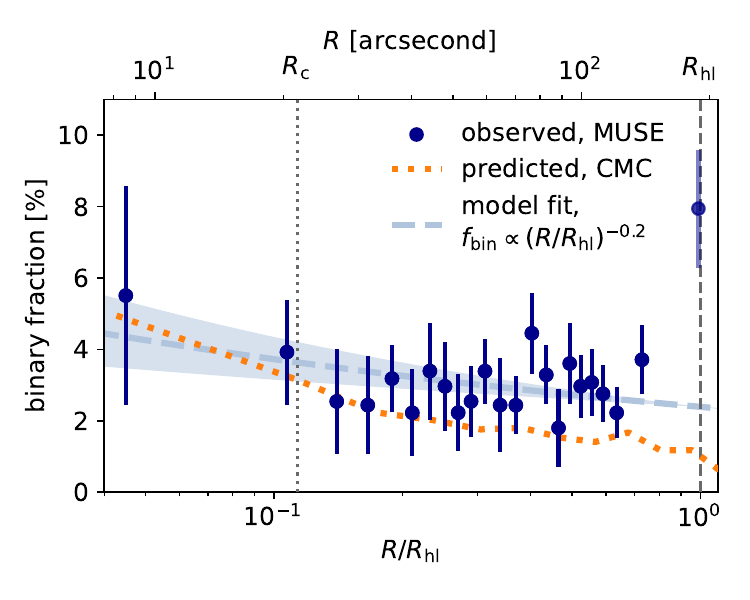}
    \caption{Binary fraction in 47\,Tuc as a function of (projected) distance from the cluster center, in units of half light radii $R_{\mathrm{hl}}$. The observed discovery fraction is shown in radial bins of 1,000 stars (dark blue), with error bars representing uncertainties, calculated by varying the binary probability threshold  $0.4 < P(\chi^2) < 0.6$ and adding statistical noise in quadrature.  The predicted binary fraction from the CMC simulation is shown as the orange dotted line. The grey dashed line represents a fitted power law, $f_{\mathrm{bin}} \propto (\nicefrac{R}{R_{\mathrm{hl}}})^{\beta}$, with $\beta = (-0.2 \pm 0.07)$; shaded areas show the uncertainty intervals.}
    \label{fig:binary_fraction_radial_profile}
\end{figure}

The available MUSE data set offers the unique opportunity for an independent estimate of the radial profile of the binary fraction, that is not limited to MS-MS binaries. In Fig.~\ref{fig:binary_fraction_radial_profile}, the observed binary discovery fraction is plotted as a function of the projected distance to the center of 47\,Tuc at (Ra,~Decl) = (00h~24m~05.67s,~-72d~04m~52.6s). It can be seen that the binary fraction increases slightly towards the center of the cluster, consistent with the results of \cite{Ji&Bregman2015}. Since only the projected distances are available, the observations are biased to somewhat dampen any trend, and the actual radial increase is likely somewhat steeper. The observed radial trend in the binary fraction aligns qualitatively with the CMC simulation, as depicted in Fig.~\ref{fig:binary_fraction_radial_profile}. Although the binary fraction in the CMC simulation is somewhat lower, it also displays little radial dependence and only a moderate increase toward the cluster core. The apparent rise in the binary fraction at larger radii, especially around the half-light radius, is probably spurious. This is because these distances are covered by only two pointings, making the observations less reliable with lower angular completeness for radii $\gtrsim 100''$. 

The radial binary fraction profile observed in 47\,Tuc is well represented by a power law function, $f_{\mathrm{bin}} \propto (\nicefrac{R}{R_{\mathrm{hl}}})^{\beta}$, where the exponent $\beta$ is determined to be $\beta = (-0.2 \pm 0.07)$, as shown in Fig.~\ref{fig:binary_fraction_radial_profile}. Notably, this exponent closely aligns with the value of $\beta = -0.23$ reported by \cite{Aros+2021} for GCs harboring a central intermediate-mass black hole (IMBH), as opposed to those without an IMBH, where $\beta = -0.33$. However, these results from \cite{Aros+2021}, derived from MOCCA simulations, appear to conflict with the findings of CMC simulations \citep{Ye+2022}, which lack a central IMBH but exhibit a flat radial profile for the binary component. The presence of an IMBH in 47\,Tuc has previously been suggested based on measured pulsar accelerations \citep{Kiziltan+2017}. However, recent studies propose that a system comprising stellar-mass black holes could sufficiently account for pulsar accelerations, thereby rendering the presence of an IMBH unnecessary \citep{Mann+2019, Ye+2022}.


\subsection{Global binary fraction of 47\,Tuc}
\label{sec:binary_fraction_total}


To overcome observational limitations, namely the limited detection efficiency, the restriction to the MUSE FoV, and the lower magnitude limit of the instrument, and to estimate the true binary fraction of 47\,Tuc, the dedicated CMC mock data set was used. 

Binary candidates discovered within the MUSE FoV will always comprise both true positives (TP) and false positives (FP). The discovery binary fraction is calculated as the sum of these binaries divided by the total number of stars in the sample. To derive the true binary fraction, encompassing both TP and false negatives (FN), we need to multiply the discovery binary fraction by the factor (TP+FN)/(TP+FP). Although exact counts of TP, FP, and FN binaries are unknown, we approximated this scaling factor through injection and retrieval tests using the CMC mock data set. Adopting a variability probability threshold of $P(\chi^2)>0.5$ as a compromise between the TP and FP rates, and varying it between 0.4 and 0.6 to estimate uncertainties, we inferred a scaling factor of $1.09 \pm 0.21$. Consequently, the observed discovery fraction of $f_{\mathrm{bin, discovery}} = (3.3\pm1.1)\%$ translates to a corrected binary fraction of $f_{\mathrm{bin, FoV}} = (3.6\pm1.4)\%$ within the FoV. 

The correction applied to the binary fraction involves implicit assumptions about the properties of the binaries that can affect the TP/FN rate. We estimated the magnitude of this effect by varying the orbital parameters and RV uncertainties of the simulated binaries within reasonable bounds, and we observe that the corrected binary fractions generally fall within the quoted uncertainties.

The procedure described above allows the underlying binary fraction in 47\,Tuc to be determined, but the estimate is restricted to the MUSE FoV and to the observable stars above the lower magnitude limit. 
Estimating the total binary fraction of 47\,Tuc, requires assumptions about how the binary fraction scales with the magnitude of the primary star and about the radial distribution of binaries in the cluster.
If we assume that the CMC simulation realistically portrays the distribution of binaries in the cluster, we can compare the total binary fraction of the simulation (1.73\%) to the binary fraction in the mock observations of detectable stars in the FoV ($2.64 \pm 0.04 \%$), which yields a ratio of $0.66 \pm 0.01$ of these two values. By multiplying this ratio with the corrected MUSE binary fraction in the FoV, the discovery fraction can be translated to a total binary fraction of $$f_{\mathrm{bin, total}} = (2.4\pm1.0)\%, $$ always under the assumption that the distribution of binaries in the CMC simulations is comparable to that of the real binary population.
The somewhat lower value of the estimated total binary fraction compared to the observed value is due to a slight increase of the binary fraction towards the cluster core due to mass segregation.

\section{Binary population demographics in 47\,Tuc}
\label{sec:binary_population}

\subsection{Determination of orbital parameters}
\label{sec:orbital_parameter_determination}

Among the 708 binary candidates identified in the previous chapter via individual variability probabilities, there are 227 binary candidates with a minimum of ten epochs. For these stars, orbital parameters were inferred using the UltraNest nested sampling framework with priors and settings as described in Sect.~\ref{sec:nested_sampling}. The sampling of the orbital parameters was limited to stars with at least ten observations since for fewer data points the posterior PDFs are generally highly multi-modal and do not yield conclusive results. 

Even for stars with ten or more epochs the posterior samples are in many cases not well constrained in this sense, and their posterior PDFs cannot be adequately described by compressing them to a single median or MAP value. We, therefore, chose to construct a catalog containing only the subsample of binaries with converged posterior parameters. This catalog comprises the subset of binaries that meet the following more stringent criteria:

\begin{enumerate}
    \item[a)] Binaries were selected that have well-constrained posterior period samples returned by UltraNest, that is they have uni- or bimodal distributions as defined by \cite{Giesers+2019} (see also Sect.~\ref{sec:nested_sampling}). This applies to 42 binaries.
    \item[b)] As a cross-check of the results, the orbital parameters were determined also via rejection sampling with \href{https://thejoker.readthedocs.io/en/latest/}{The Joker} \citep{Price-Whelan+2017}. Only the 34 binaries that are well-constrained by both methods were retained in the subset.
    \item[c)] In order to test the robustness of the sampling, the posterior was re-evaluated with UltraNest for the RVs determined via the iterative cross-correlation method (see \cite{Saracino+2023b} for a detailed description). Binaries that have inconsistent posterior periods are removed. This leaves 30 binaries in the final subset, of which 23 are uni- and seven bimodal. The binary systems are spread accross the FoV and have their properties summarized in Table~\ref{tab:binary_system_properties}.
\end{enumerate}

Step b), the cross-check of the results with The Joker, is to ensure that the results found are independent of the sampling method. Eight binary stars were removed, for which rejection sampling with The Joker did not yield well-constrained orbits. Step c) reduces the risk of finding spurious highly eccentric orbits caused by single outliers in the RVs that remained in the sample despite the rigorous filtering process.

\subsection{Period distribution}
\label{sec:period_distribution}

\begin{figure*}[t]
\centering
\begin{subfigure}{.75\textwidth}
    \centering
    \includegraphics[width=\textwidth]{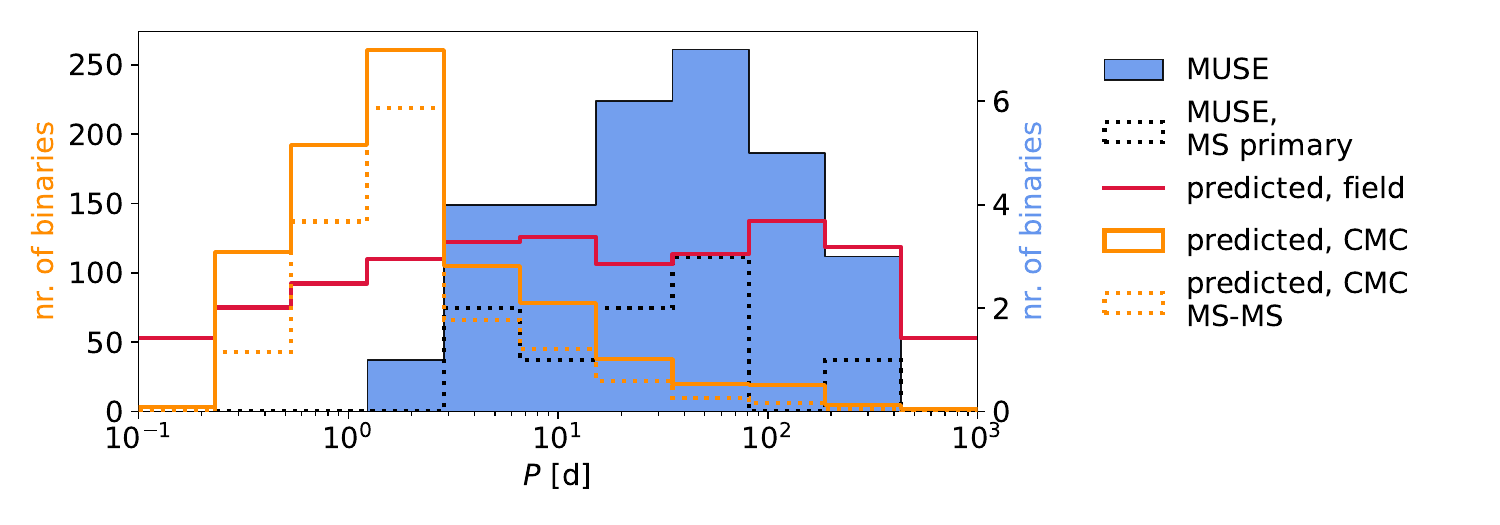} 
\end{subfigure}%
\begin{subfigure}{.25\textwidth}
    \centering
    \includegraphics[width=\textwidth]{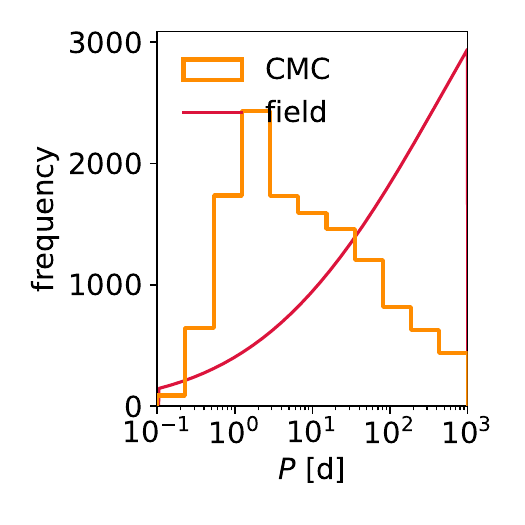} 
\end{subfigure}
\caption{Period distribution of binaries with well-constrained orbits in 47\,Tuc (shown in blue). The dotted line represents the subset of binaries with MS primaries. For comparison, the orange lines depict the period distributions of hypothetically detectable binaries from the CMC simulation (stellar types as shown in Fig.~\ref{fig:ktype_heatmap}), with the dashed line indicating the subset of simulated MS-MS binaries. The red curve illustrates the predicted observable distribution assuming an underlying field-like period distribution with $\overline{\log P} \approx 5.0$ and standard deviation $\sigma_{\log P} \approx 2.3$. It is scaled to the number of well-constrained MUSE binaries. For CMC and field binaries, we have forward-modeled our selection function to account for decreasing sensitivity at longer periods. The underlying period distributions of field and CMC binaries are shown for reference in the right-hand panel.}
    \label{fig:period_distribution}
\end{figure*}

We present the orbital period distribution of well-constrained binaries in Fig.~\ref{fig:period_distribution}. The inferred periods span a range from 2.7 days to 425 days and are well described by a log-normal distribution with mean $\overline{\log P} = 1.5$ and standard deviation $\sigma_{\log P} = 0.6$ for $P$ in days. We compare the observed distribution of periods to the distributions of binaries in the Galactic field and the simulated binaries in the CMC simulation of 47\,Tuc. 

The period distribution of solar-type stars in the field follows a log-normal distribution with mean $\overline{\log P} \approx 5.0$ and standard deviation $\sigma_{\log P} \approx 2.3$ \citep{Raghavan+2010, Moe&DiStefano2017}, peaking at periods of several years. Its general shape is plotted in the right-hand panel of Fig.~\ref{fig:period_distribution}, for reference. In the main panel of Fig.~\ref{fig:period_distribution}, we show the expected observed period distribution for field-like binaries after forward modeling our selection function. The apparent bias toward short periods is a consequence of the MUSE sensitivity, which is inversely correlated with the orbital period (see Sect.~\ref{sec:sensitivity}). Comparison with the MUSE sample reveals a deficit of long-period binaries ($\sim \mathcal{O}(10^3)$ days) in 47\,Tuc compared to the field. This is not surprising given the expected depletion of binaries beyond the boundary between hard and soft orbits, roughly at a few hundred days in the case of 47\,Tuc, depending on the masses of the binary components. Soft binaries (i.e., binaries with an orbital energy below the average kinetic energy in the cluster) are easily disrupted during dynamical encounters, quickly depleting the cluster of long-period binaries. 

The CMC simulation generates binaries with a high number of close binary systems. Consistent with the predictions of \cite{Ivanova+2005}, the simulations exhibit a period distribution centered on periods of only a few days. This characteristic holds for both the hypothetically detectable subset and the entire binary population (see right panel of Fig.~\ref{fig:period_distribution}), indicating that the shape is not merely a selection effect. Interestingly, the period distribution observed in the MUSE binary systems differs notably from the predictions of the CMC simulations, with the MUSE binaries peaking at periods of weeks to months instead of days and lacking binaries with periods $\lesssim 1\,$day. This highlights a significant deficit in short-period binaries in the observed MUSE sample compared to the simulations, which is particularly notable considering the increased observational sensitivity to short periods. There is also an apparent deficit of short-period binaries in 47\,Tuc when compared to the field population, suggesting that the frequency of short-period binaries in 47\,Tuc may even be lower than in the Galactic field.

\subsection{Eccentricity distribution}
\label{sec:eccentricity_distribution}
\begin{figure}[ht!]
    \centering
    \includegraphics[width=\columnwidth]{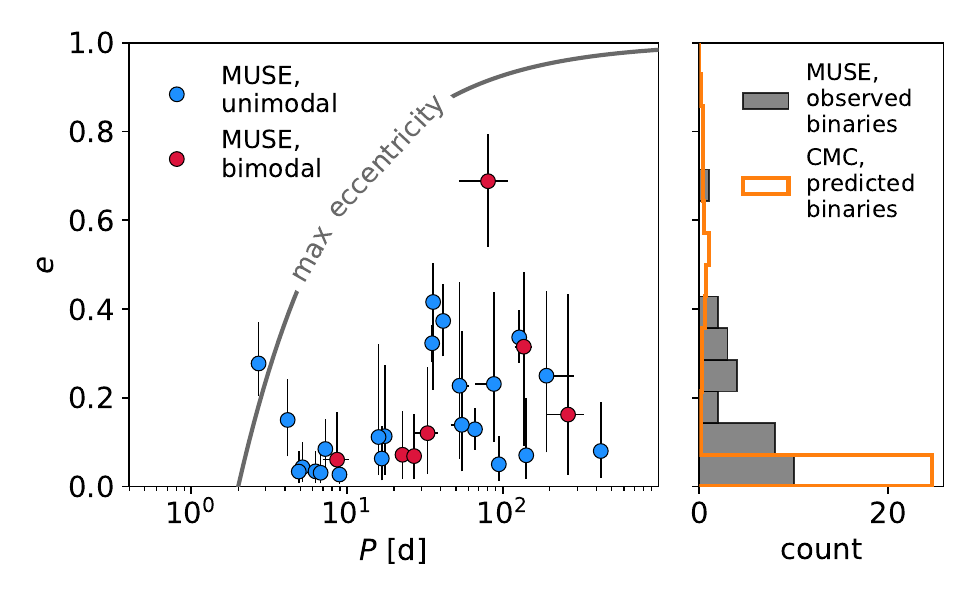}
    \caption{Period-eccentricity distribution of the 30 well-constrained binaries with uni- (blue) or bimodal (red) posterior period distributions. The maximum eccentricity limit as defined by \cite{Moe&DiStefano2017} is indicated as a solid gray line. The panel on the right-hand side shows a collapsed histogram of the eccentricity distribution for the well-constrained MUSE binaries. For comparison, the predicted eccentricity distribution of the detectable CMC binaries is plotted in orange, scaled to the number of MUSE binaries for better readability.}
    \label{fig:eccentricity}
\end{figure}

In Fig.~\ref{fig:eccentricity}, the period eccentricity distribution of the well-constrained binaries is shown. The measured eccentricities have values between zero and 0.7. The solid gray line indicates the maximum eccentricity limit for MS stars. It is defined as the limit below which MS binaries have Roche Lobe fill factors of $\leq 70\%$ at periastron, following \cite{Moe&DiStefano2017}. Except for one system, all binaries have eccentricities well below this limit; $e \ll e_{\mathrm{max}}$. Even the shortest-period system close to the eccentricity limit is within the uncertainties still consistent with a detached, non-Roche lobe filling binary. We note that RV observations are generally biased towards lower eccentricities since circular orbits require on average fewer data points to be uniquely identified.
From the eccentricity distribution, it appears as if the observed eccentricities are slightly higher than those in the simulations. This could be correlated with the longer periods, for which tidal circularization of the orbits is less effective. However, eccentricity posteriors are associated with larger uncertainties. Therefore, the binaries with eccentricities up to ca. 0.3 are mostly consistent within their uncertainty with circular orbits. 

\subsection{Companion masses and mass ratio distribution}
\label{sec:companion_mass}

\begin{figure}[t!]
    \centering
    \includegraphics[width=0.9\columnwidth]{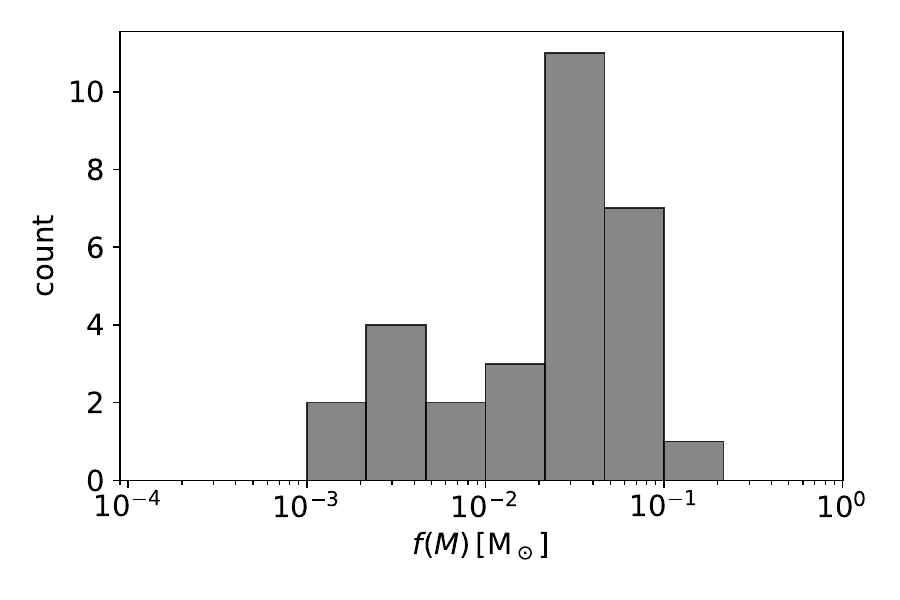}
    \caption{Binary mass function $f(M)$ for the 30 well-constrained binaries in 47\,Tuc. All binaries are characterized by $f(M) \ll 1\,\mathrm{M}_\odot$.}
    \label{fig:binary_mass_function}
\end{figure}

For well-constrained binaries, the binary mass function $f(M)$ was directly inferred from the posterior parameters returned by UltraNest, without the need for assumptions about the mass of the observed primary star. The mass function is defined as
\begin{equation}
    f(M) \equiv \frac{(m_2 \sin{i})^3}{(m_1+m_2)^2} = \frac{1}{2\pi G}\frac{K_1^3}{f} (1-e^2)^{\frac{3}{2}}\,.
\end{equation} 
The resulting distribution of binary mass functions is plotted in Fig.~\ref{fig:binary_mass_function}, showing that all observed binaries exhibit low mass functions $f(M) < 0.2\,\mathrm{M}_\odot$. The distribution of the mass function is broadly consistent with the CMC simulations, except the CMC simulations predict a tail of high $f(M) \gtrsim 0.25\,\mathrm{M}_\odot$ binaries with massive white dwarf, neutron star, and BH companions. The absence of such binaries in the MUSE sample makes the presence of dark companions among the observed binaries unlikely. 

\begin{figure*}[t!]
    \centering
    \begin{minipage}[t]{0.5\textwidth}
        \centering
        \includegraphics[width=\textwidth]{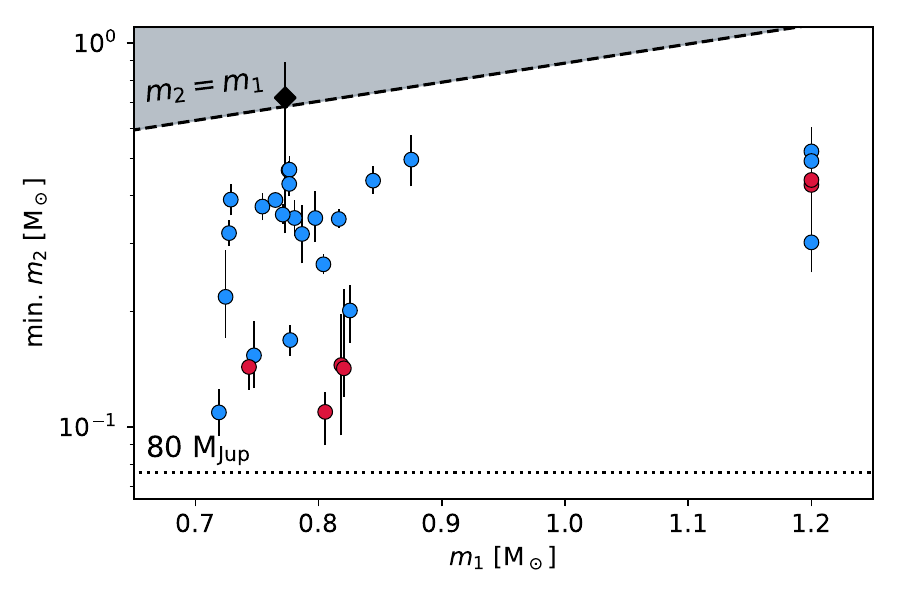}
    \end{minipage}%
    \begin{minipage}[t]{0.5\textwidth}
        \centering
        \includegraphics[width=\textwidth]{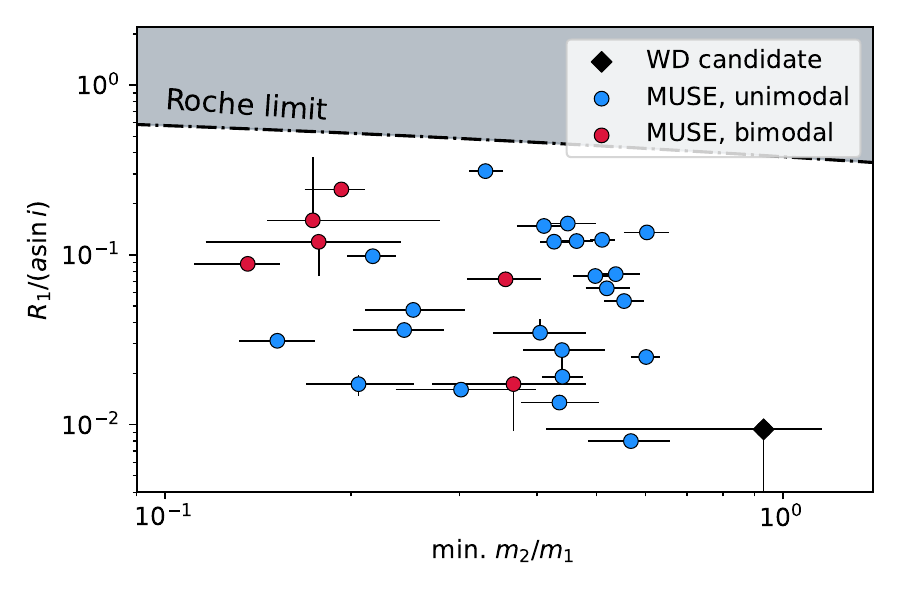}
    \end{minipage}
    \caption{Inferred minimum companion masses and mass ratios of binaries with well-constrained orbits. {Left:} Minimum companion masses $m_2$ plotted against primary mass $m_1$ for the reduced subset of binaries with uni- (blue) and bimodal (red) posterior solutions in 47\,Tuc. The dashed line indicates a mass ratio of $q=1$, the dotted line marks the approximate hydrogen burning limit. {Right:} Primary stellar radius $R_1$ divided by the determined (minimum) binary semi-major axis $a$ as a function of minimum mass ratio. The color scheme is the same as in the left-hand figure. The dash-dotted line indicates the Roche radius. The black diamond marks a binary with a possible white dwarf companion.}
    \label{fig:companion_masses}
\end{figure*}

Direct computation of the actual companion masses $m_2$ is not possible since the orbital inclinations are unknown. However, assuming an inclination of $90^{\circ}$, corresponding to an edge-on orbit, and using estimates of primary masses from isochrone fits, the minimum companion masses can be inferred numerically. Figure~\ref{fig:companion_masses}, inspired by Fig.~9 in \cite{Price-Whelan+2020}, shows these minimum companion masses relative to the primary masses for the well-constrained binaries. The dashed line indicates the mass equality limit, while the dotted line represents the hydrogen burning limit. 29 out of 30 well-constrained binaries fall between these two limits, suggesting they are not brown dwarf candidates below the hydrogen burning limit or dark compact object candidates near or above the mass equality limit. One system, indicated in black, has an estimated mass ratio close to unity and may feature a white dwarf companion. It is described in more detail in Sect.~\ref{sec:peculiar_objects}.

The right panel of Fig.~\ref{fig:companion_masses} shows the ratio of the primary stellar radius $R_1$ over the (projected) system semi-major axis $a \sin(i)$ as a function of the minimum mass ratio. The approximate Roche limit, as parameterized by \cite{Eggleton1983} is indicated with a dashed line. All well-constrained binary systems are well within their respective Roche limits. We note that only the minimum semi-major axes can be determined, so the quoted values of $\nicefrac{R_1}{a \sin(i)}$ are upper limits and will be smaller for low inclinations. 

\begin{figure}[t!]
    \centering
    \includegraphics[width=\columnwidth]{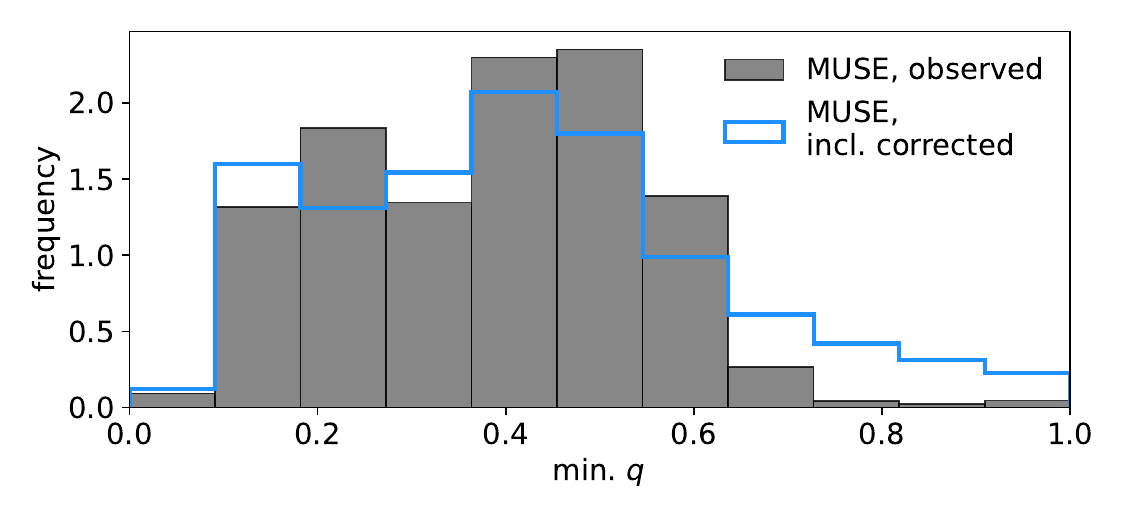}
    \caption{Normalized mass ratio distribution of the subset of well-constrained binaries in 47\,Tuc. The gray filled histogram shows the distribution of minimum mass ratios derived from the UltraNest posterior samples. The blue line shows the same results but with  a statistical correction to account for orbital inclinations. 
    }
    \label{fig:mass_distribution}
\end{figure}

A histogram of the mass ratio $q$ is shown in Fig.~\ref{fig:mass_distribution} for well-constrained binary systems. For each binary system, the (minimum) mass ratio was resampled 1000 times from the posterior distribution to account for the uncertainties in the orbital parameters. In addition, an inclination-corrected version of the mass ratio distribution is shown, where the inclinations were drawn from a uniform distribution in $\cos(i)$. 
The observed mass ratio distribution (inclination corrected) appears to be nearly flat for low values of $q \leq 0.5$, consistent with the results of \cite{Milone+2012} and the setup of the CMC simulations. However, for high values of $q$, a significant decrease in the number of binaries is observed. This decrease could indicate a nonuniform distribution of the mass ratio, consistent with models in which binary stars form as random pairs of stars, and in line with the results of \cite{Ji&Bregman2015}. We point out that the observed mass ratio distribution has not been corrected for mass ratio-dependent sensitivity, specifically damping of RV variability for similarly bright MS binaries (see Sect.~\ref{sec:CMC_mock_data}). We opted against such a correction, as determining sensitivity corrections based on the mass ratio (as discussed in Sect.~\ref{sec:sensitivity}) is prone to fail when addressing systems entirely missing from the sample (i.e., MS binaries with high mass ratios). Consequently, the inclination-corrected mass ratio distribution may still exhibit a bias towards lower values of $q$, preventing definitive conclusions about its overall shape.

\subsection{Peculiar objects}
\label{sec:peculiar_objects}
\begin{figure*}[t!]
    \centering
    \includegraphics[width=0.7\textwidth]{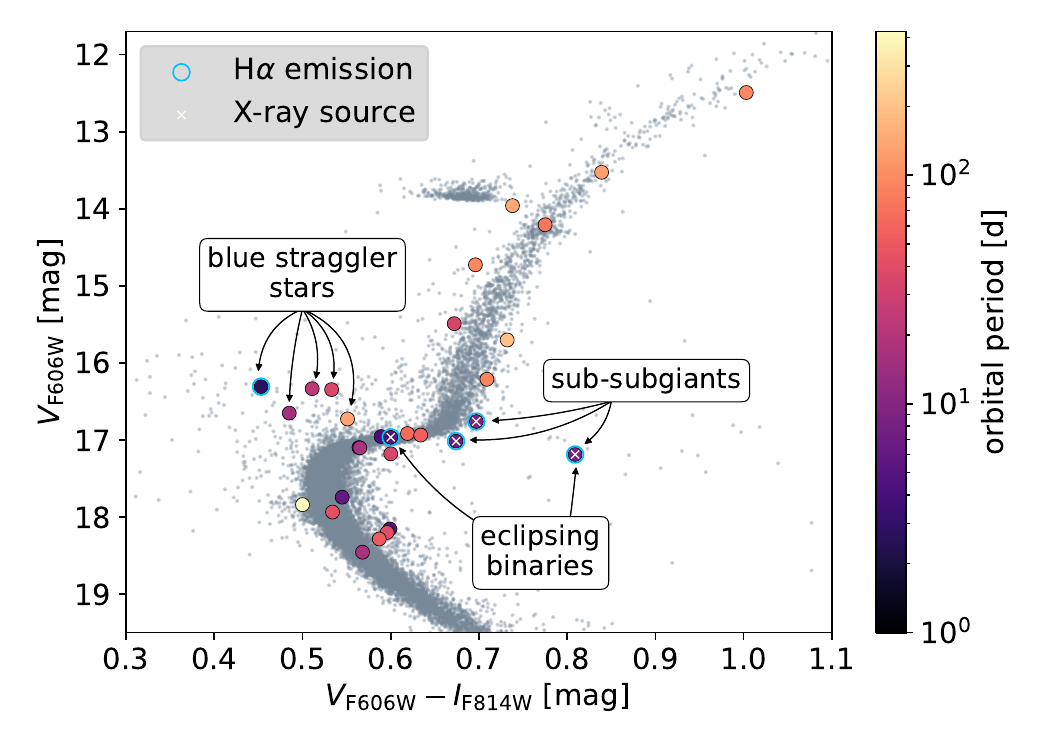}
    \caption{Color-magnitude diagram of 47\,Tuc. The binaries with well-constrained orbital parameters are highlighted and color-coded according to their orbital periods. Two known eclipsing binaries, five blue straggler stars and three sub-subgiants are indicated with arrows. Emisson line objects and possible X-ray sources are marked with blue circles and white crosses, respectively.}
    \label{fig:CMD_highlighted}
\end{figure*}

GCs are homes to a large number of stellar exotica, such as BSS and sub-subgiants. Their formation is likely linked to dynamic interactions in binary systems and facilitated by the high stellar densities and resulting elevated encounter rates in GCs. Binaries with well-constrained orbits allow us to probe these populations of peculiar stellar objects, including the search for dark remnant companions. Figure~\ref{fig:CMD_highlighted} shows the positions of the well-constrained binaries in the CMD of 47\,Tuc. The binaries span a broad range of primary stellar types, from MS to RGB stars.

\subsubsection{Blue straggler stars}
The MUSE data set contains 67 BSS with at least ten epochs, of which seven are likely to belong to binary systems. For five BSS, UltraNest provides well-constrained orbital solutions with periods ranging from 2.7 to 133.7\,d and minimum companion masses between 0.25 and $0.43\,\mathrm{M}_\odot$. \cite{Giesers+2019} identified a bimodal distribution of binary periods of BSS in the GC NGC\,3201 with, on the one hand, very hard binaries ($P \lesssim 1\,$d) and, on the other hand, relatively wide binaries ($P > 100\,$d). 
Interestingly, the BSS binaries observed in 47\,Tuc do not show signs of two distinct populations with four of five BSS systems between the two regimes of periods mentioned above. However, we note that there exist observations of some short-period BSS in eclipsing binaries outside the MUSE FoV \citep{Weldrake+2004}. It is conceivable that the differing period distributions of BSS binaries in the two GCs are a consequence of different BSS formation histories and that BSS properties vary with cluster environment. Differences between the BSS populations of the two GCs have already been observed in terms of different rotational velocities of the BSS \citep{Ferraro+2023}, for example. 

\subsubsection{Sub-subgiant stars}
Sub-subgiant (SSG) stars are classified as stars that are optically redder than (binary) MS stars and fainter than regular subgiants. Their position in the CMD is not easily explained by standard single-star stellar evolution theory. Possible mechanisms for their formation include mass loss in a mass-transferring binary, stripping of a subgiant’s envelope and reduced luminosity due to magnetic fields \citep{Leiner+2017}. By studying the demographics of SSGs in star clusters, \cite{Geller+2017} identified $\sim 50$\% of the sample as RV variable with typical periods less than 15\,d, 58\% of the SSGs were found to be X-ray sources and around a third showed H$\alpha$ emission.

Among the well-constrained binaries in 47\,Tuc, one clear SSG and two candidates can be identified. The stars have orbital periods of 6.77, 8.47 and 8.94\,d, consistent with the typical upper limit identified by \cite{Geller+2017} and circular orbits. All three stars could be matched to X-ray sources from the \textit{Chandra} Source Catalog \citep{Evans+2010} with source ids 2CXO J002358.7-720447, J002413.7-720334, and J002417.2-720301. Furthermore, all three stars exhibit variable H$\alpha$ emission, which shows in the form of partially filled H$\alpha$ lines with respect to the model spectra. X-ray and H$\alpha$ emission can be interpreted as signs of chromospheric and surface magnetic activity \citep{Leiner+2017} or may be indicative of ongoing mass transfer. 

\subsubsection{Tip of the RGB stars}
Among the well-constrained binary systems, there is one system that has a very bright primary star ($M_\mathrm{F606W} = 12.49$) located near the tip of the red giant branch. The system in question has a relatively long orbital period of 94.3\,d. A period near 100\,d is suspicious for an RGB binary candidate, as these periods are typical pulsation periods of long-period variables. Therefore, it seems likely that the binary star is a false positive resulting from asteroseismic modes manifesting as RV jitter. The primary star is neither listed as a variable in the catalog by \citet{Clement+2001} nor in the Hubble Catalog of Variables \citep{Bonanos+2019}. Additionally, no significant photometric variability was detected in the MUSE data, which explains why the star remained in our sample despite the cuts applied for photometric variability (see Sect.~\ref{sec:photometric_variables}).

\subsubsection{Eclipsing binaries}
One SGB star and one SSG star in the sample are of particular interest, as they are already cataloged binary stars. The SSG with the identifier WF2-V32 has been detected as a variable red straggler by \cite{Albrow+2001} and is listed as an eclipsing binary in the catalog of variable stars by \cite{Clement+2001}. Given the comparably low number of ten available observational epochs, the orbital constraints returned by UltraNest for this star are not fully converged but bimodal, with peaks at approximately 3.6 and 8.5\,d. Taking into account the period of 9.2\,d estimated by \cite{Albrow+2001}, albeit from saturated photometry, the lower period found by UltraNest seems less likely. 

The SGB star with the identifier WF4-V04 has been classified as an eclipsing Algol-type binary with a period of 4.92\,d \citep{Albrow+2001}. It exhibits an excess in H$\alpha$ and can be matched to a nearby \textit{Chandra} X-ray source. The posterior distribution returned by UltraNest precisely confirms the literature orbital period. 
The fact that these two binaries are eclipsing implies that the orbits are observed nearly edge-on. Therefore, the minimum companion masses determined for $0.14\,\mathrm{M}_\odot$ and $0.47\,\mathrm{M}_\odot$ for assumed primary masses of $0.74\,\mathrm{M}_\odot$ and $0.78\,\mathrm{M}_\odot$ will be approximately the same as the true companion masses. However, it should be noted that the primary mass of the SSG determined from an isochrone fit is associated with large uncertainties.

\subsubsection{White dwarfs}

White dwarfs are relatively faint, and no white dwarf primaries are contained in the filtered MUSE data set. Nevertheless, they should be detectable in binaries via RV variability of their luminous companions. The more massive white dwarfs can in principle be identified as binaries with moderately high companion masses and high mass ratios $q$ close to unity. Among the well-constrained binaries, there is one candidate system with a possible white dwarf companion. It has a SGB primary star of approximately $0.77\,\mathrm{M}_\odot$ and is marked in black in Fig.~\ref{fig:companion_masses}. The system has a bimodal orbital solution returned by UltraNest, with posterior samples clustered around orbital periods of $\sim 70$ and $\sim 280 $\,d, giving minimum companion masses of $\sim 0.3$ and $\sim 0.8\,\mathrm{M}_\odot$, respectively. A companion mass of $\sim 0.8\,\mathrm{M}_\odot$ would imply a mass ratio greater than one, indicating the presence of a faint stellar remnant companion. More RV observations are needed to determine the true orbital period and resolve the ambiguity. 

Using the mass function only allows us to detect white dwarf companions with masses above that of the visible companion. Since RV precision is decreasing towards late spectral types on the MS, the least massive visible companion has a mass of about 0.6\,$\mathrm{M}_\odot$. This approach is therefore limited to the detection of white dwarfs with masses exceeding those of typical white dwarfs. For binaries with a MS primary as the visible component, the secondary can either be a less luminous and therefore less massive MS star or a compact object, given the mass range in question most likely a white dwarf. Although the compact object effectively does not contribute to the spectrum, a MS companion can, depending on the flux ratio. As mentioned in Sect.\,\ref{sec:CMC_mock_data}, this leads to an attenuation of the observed RV amplitude in our low-resolution spectra. Varying the companion mass $m_2$ and the inclination $i$ in Eq.\,\ref{eq:RVamplitude} and keeping the primary mass $m_1$ within its uncertainty, combinations of $m_2$ and $i$ can be identified that match the observed RV amplitude $K_1$. Assuming a MS secondary, the flux ratio can be estimated from the given masses of the two companions, and a corresponding attenuation factor can be applied. For a given $i$, $K_1$ therefore first increases with increasing $m_2$ but then reaches a maximum and decreases when the flux ratio reduces. For equal mass binaries, the RV amplitude is then zero. Therefore, it is possible that no combination of $m_2$ and $i$ reaches the observed RV amplitude for a MS pair. In those cases, the secondary has to be a compact object, not affected by the attenuation. This also allows us to distinguish MS from white dwarf companions for white dwarfs with masses lower than that of the primary. We applied this method to binaries with MS primaries and found none where the observed RV amplitude could not be explained with a MS pair.   

\subsubsection{Neutron stars and BHs}
From the observed stars in the MUSE data set, no evidence for neutron star or BH companions is found. In principle it is possible that there are massive remnants among the identified companion stars if their higher masses are disguised by low inclinations. We estimated this probability from the inclination-corrected distribution of companion masses in a statistical approach, as was done for the mass ratio distribution. Only in 4.2\% of the cases are the determined companion masses above the Chandresekhar limit of $1.4\,\mathrm{M}_\odot$, indicating a neutron star, and in less than 2.6\% of the cases are the companion stars more massive than $2.1\,\mathrm{M}_\odot$, the approximate theoretical upper limit for neutron stars and the expected lower limit for BH masses. This implies that while there is a small chance of massive dark remnants among the observed binary stars, the probabilities are tiny and would require a nearly face-on orbital configuration.

\section{Discussion}
\label{sec:discussion}

\subsection{Observed spectroscopic binary fraction in the context of photometric estimates}
\label{sec:literature_comparison}
The binary fraction of 47\,Tuc has previously been estimated by \cite{Milone+2012} and \cite{Ji&Bregman2015} using photometry from HST WFPC2 and ACS observations. Interestingly, the total binary fraction of $(2.4\pm1.0)\%$ determined in our study with MUSE falls within the range of photometrically determined values of $(1.8\pm0.6)\%$ \citep{Milone+2012} and $(3.01\pm0.13)\%$ for $q>0.5$ \citep{Ji&Bregman2015}.

However, directly comparing MUSE and photometric estimates is not trivial because of differing observational biases. With MUSE, we focus on brighter stars, reaching approximately one to two magnitudes below the MS turn-off, and apply a correction for lower magnitudes. In contrast, \cite{Milone+2012} estimate the binary fraction solely for MS stars in the F814W magnitude interval ranging from 0.75 to 3.75 magnitudes below the MS turn-off, causing potential discrepancies if the binary fraction of MS-MS stars differs from the general binary fraction, although the MUSE data reveals no strong dependence of the binary star fraction on stellar type. 
In addition, the photometric binary fraction remains uncorrected for binaries with dark companions, like MS-white dwarf binaries, which cannot be detected. This inherent limitation in the photometric approach may result in an underestimation of the total binary fraction.

In addition, the studies focus on different regions of the cluster. The MUSE FoV covers the center of the GC out to the half-light radius, whereas the photometric study by \cite{Milone+2012} is limited to radii between the core and the half-mass radius of the cluster, and the region covered by \cite{Ji&Bregman2015} includes the core but extends only to 0.75 half-light radii. This implies that the central increase and outer decrease in binary fraction are not covered by the photometric data of \cite{Milone+2012} and \cite{Ji&Bregman2015}, respectively. 

Moreover, photometric estimates are most sensitive to high mass ratios $q>0.5$ and the results need to be extrapolated for lower values of $q$. 
In contrast, the sensitivity of the RV method has limited sensitivity for high mass ratios close to one. 
If the distribution of $q$ is slightly skewed toward low mass ratios instead of flat - as suggested by \cite{Ji&Bregman2015} - the extrapolation of the photometric estimates would underestimate the true binary fraction. 

\subsection{The binary fraction landscape of massive star clusters}
\label{sec:cluster_comparison}
\begin{figure}[t]
    \centering
    \includegraphics[width=\columnwidth]{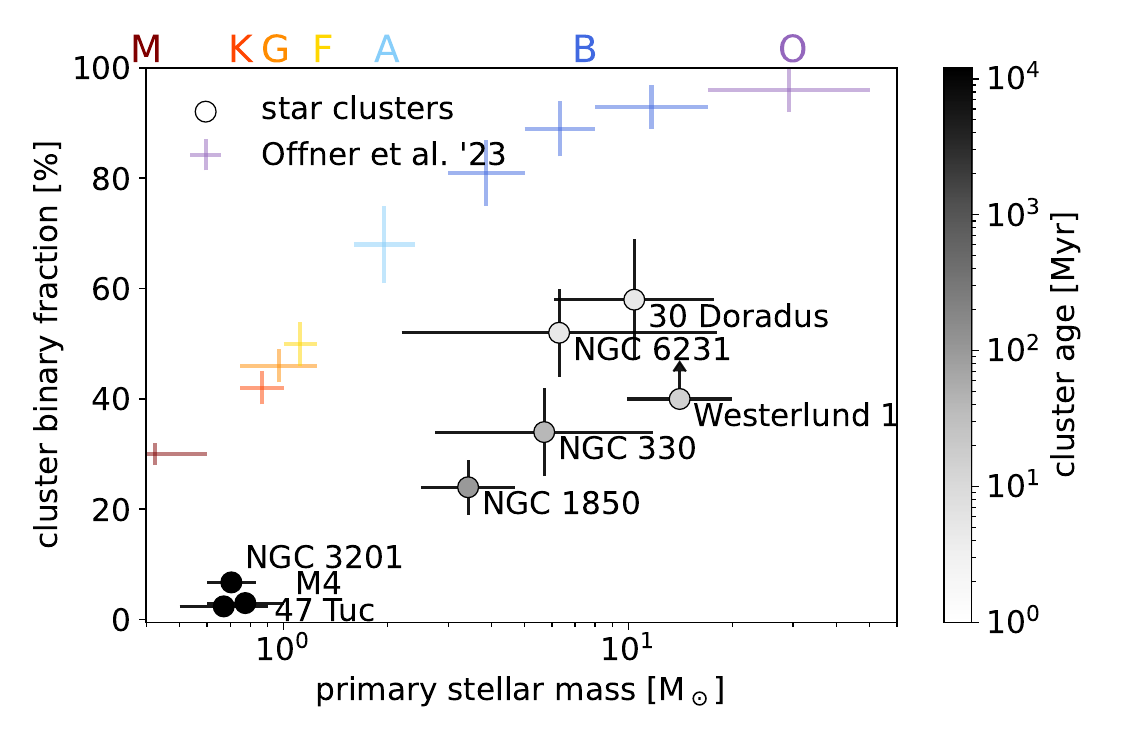}
    \caption{Spectroscopic binary fractions and respective uncertainties of massive star clusters as a function of primary stellar masses. Horizontal bars indicate the approximate mass ranges probed by the respective studies estimated based on the indicated spectral types, see Sect.~\ref{sec:cluster_comparison} for references. The grey-scale color gradient shows approximate cluster ages. For comparison, the multiplicity statistics for different spectral types and masses from \cite{Offner+2023} are shown as crosses (see their Table~1).}
    \label{fig:cluster_comparison}
\end{figure}

We compare the determined close binary fraction of 47\,Tuc with the spectroscopic binary fractions reported for other massive star clusters in Fig.~\ref{fig:cluster_comparison}. These clusters cover a range of ages, from a few Myr for young clusters such as 30\,Doradus, NGC\,6231, and Westerlund\,1, to intermediate ages of 40 to 100, represented by clusters such as NGC\,330 and NGC\,1850 and to old clusters more similar to 47\,Tuc, such as NGC\,3201 and M\,4, with ages $\gtrsim 10\,$Gyr.

Figure~\ref{fig:cluster_comparison} shows a discernible trend of the binary fraction with cluster age, with older clusters exhibiting lower binary fractions. For example, binary fractions exceeding $50\%$ are reported in 30\,Doradus \citep{Dunstall+2015} and NGC\,6231 \citep{Banyard+2022}, a lower limit of $40\%$ in Westerlund\,1 \citep{Ritchie+2022}, and between $(24\pm5)\%$ and $(34\pm8)\%$ in NGC\,1850 \citep{Saracino+2022} and NGC\,330 \citep{Bodensteiner+2021}, whereas NGC\,3201, M\,4, and 47\,Tuc all show binary fractions of less than 10\% \citep{Giesers+2019, Sommariva+2009}. This apparent age dependence likely arises from the varying stellar masses present in the clusters; with younger clusters still retaining a substantial fraction of O- and early B-type stars. However, it should be noted that comparing the binary fractions of the clusters with the general multiplicity statistics of MS stars in the Galactic field and open clusters (taken from the review by \cite{Offner+2023}), consistently shows lower binary fractions in the star clusters. The low binary fractions in GCs are probably not primordial, but a consequence of the dynamical evolution of the cluster with many binaries disrupted by strong encounters with other stars or binaries \citep{Heggie1975,Cote+1996} or as a result of stellar evolution processes \citep{Fregeau+2003}. 

Among the old GCs, 47\,Tuc stands out as the most massive and densest \citep{Harris2010}, yet it exhibits the lowest binary fraction. This finding aligns with photometric studies by \cite{Milone+2012}, which suggest an anti-correlation between cluster luminosity (and hence mass) and binary fraction. To probe correlations with other cluster properties, such as metallicity, comprehensive spectroscopic observations of a larger sample of clusters are needed. 

\subsection{Deficit of short-period binaries}

The binaries observed in 47\,Tuc exhibit lower RV amplitudes and longer periods than expected from the CMC simulations. 
As demonstrated for the population of simulated binary stars (see Fig.~\ref{fig:detection_fraction}), the RV detection probability is highest for binaries with short periods, as they generally have higher RV amplitudes, and decreases for longer periods. If anything, the MUSE results should be biased toward compact binary systems. The absence of such short-period compact systems among the observed binaries is, therefore, remarkable.

The observed disparities in binary periods between the MUSE data and the CMC simulations may be attributed to uncertainties in the binary evolution code used in the simulations, particularly in the treatment of the common-envelope process. The modeling of the common-envelope process is associated with large uncertainties, potentially leading to an excess of very tight binaries containing a stellar remnant. To investigate this hypothesis, we compared the distribution of orbital periods of the MUSE binaries with those of the CMC binaries composed of two MS stars. The latter subset represents binaries that have not evolved off the MS and have not been affected by uncertainties in the binary evolution process. In Fig.~\ref{fig:period_distribution}, the period distribution of MUSE binaries is shown along with that of the CMC MS-MS binaries. Isolating the MS-MS CMC binaries from the evolved binaries does indeed reduce the fraction of very compact binaries with periods below one day. However, it does not alter the overall shape of the period distribution, which remains inconsistent with the observed population.

As a GC ages, the binary orbits may experience dynamical hardening, that is, a decrease in orbital period, due to energy loss during stellar encounters \citep{Heggie1975,Hut1983}. 
However, analyzing the CMC simulations reveals that many of the simulated binaries were already initiated with short periods. The observed peak at short periods in the period distribution of the best-fit simulation snapshot is therefore not solely a consequence of dynamical hardening but is an artifact of the initial period distribution, which already favored shorter periods. The longer periods observed with MUSE indicate a need to refine the primordial period distribution in CMC simulations. 

However, it is crucial to note that we do not dismiss the presence of short-period binaries altogether. Notably, short periods have been observed, for example, for cataclysmic variables. Instead, our argument centers on the lack of such binaries in our data set, despite the observational bias that favors their detection. Considering also the successful identification of short-period binaries with MUSE in NGC 3201 and other GCs, we contend that a period distribution peaking at short ($P \approx 2\,$d) periods is inconsistent with our observations.

\subsection{Finding massive companions}
Among the binaries observed in 47\,Tuc, no evidence for dark massive companions is found. With the highest inferred companion mass being $m_2 \sin i = 0.49\,\mathrm{M}_\odot$, the presence of heavy white dwarf, neutron star, or black hole companions in the sample is extremely unlikely and would require near face-on orbital configurations. Assuming 47\,Tuc hosts a larger population of dark massive objects, as suggested by simulations \citep{Weatherford+2020,Mann+2019,Ye+2022,Henault-Brunet+2020} and the detection of multiple low-mass X-ray binaries \citep{Heinke+2005}, the question arises as to why they have not been observed in this study. Several explanations are conceivable. 

It is possible that the dataset includes binaries with massive companions whose orbital parameters were incorrectly inferred. However, this seems unlikely, as our simulations showed that binaries with massive companions tend to have high RV amplitudes, which makes them easier to fit accurately. 

\begin{figure}[t]
    \centering
    \includegraphics[width=\columnwidth]{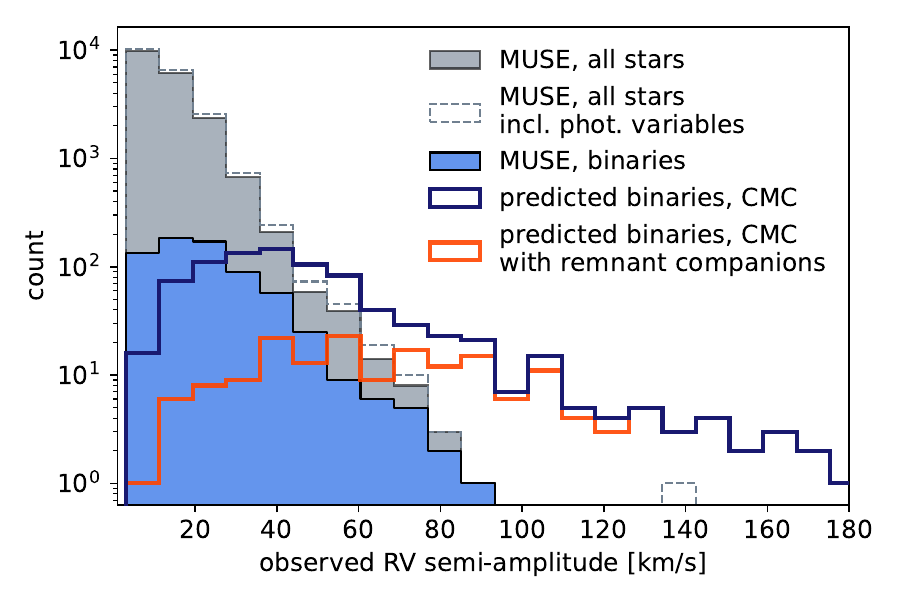}
    \caption{Observed RV semi-amplitudes, $0.5(v_\mathrm{rad,max} - v_\mathrm{rad,min})$, for all stars observed with MUSE (gray), the 708 MUSE binaries (blue), and the hypothetically observable CMC binaries from the best-fit model (dark blue line). For reference, the histogram of observed RV amplitudes is also shown for all stars without the cut on photometric variables (see Sect.~\ref{sec:photometric_variables}, dashed line), though the differences are negligible. The subset of simulated binaries with remnant (mostly white dwarf) companions is highlighted in orange. The observed lack of stars with high RV amplitudes, compared to simulations, suggests fewer stars with massive companions and/or short periods.}
    \label{fig:rv_scatter}
\end{figure}

It could also be that binary stars with massive companions have been observed but not with sufficient epochs to derive reliable orbits and companion masses. However, Fig.~\ref{fig:detection_fraction} indicates that the detection probability for binaries with massive companions is close to one, even if less than ten epochs are available. So if this were the case, the stars with massive companions should still exhibit a high amount of RV scatter, which can thus be used as a model-independent proxy for the companion masses. Figure~\ref{fig:rv_scatter} shows the distribution of the observed RV semi-amplitudes, defined as half the difference between the maximum and minimum observed RVs for each star. In the MUSE data set, all binary stars have RV scatter $< 90$km/s, which is roughly consistent with the MS-MS population of the CMC model. However, binary stars in the CMC model produce a tail of high RV scatter $> 100$km/s. This is mainly caused by MS-white dwarf binary stars. The subset of binaries with black holes or neutron stars has even higher amplitudes $\gg 100$km/s, even if they have less than ten epochs and moderate inclinations. The lack of stars with high observed RV scatter in the MUSE data implies that there are no obvious candidates for remnants in the sample. 

Another reason for the lack of massive companions could be the incompleteness of the observations due to the constrained MUSE FoV and the lower magnitude limit. However, the rapid mass segregation of massive objects to the center, also observed in the CMC simulations, makes this scenario unlikely. In addition, binary stars tend to exchange less massive companions in binary interactions \citep{Hills&Fullerton1980}. Therefore, the survival of binaries composed of massive remnants and low-mass MS stars in the dense cluster core of 47\,Tuc is less probable.

The most straightforward explanation for the lack of observed massive companions is that there simply are no or very few spectroscopically detectable binaries with massive companions in 47\,Tuc. This may indicate that the total population of remnants is small. However, the number of (accreting) BHs in binaries present at late times during GC evolution does not necessarily correlate with the total number of retained black holes \citep{Kremer+2018,Chatterjee+2017}. Alternatively, massive companions may reside in configurations that are not observable with MUSE, for instance, as single objects or in remnant-remnant binaries like the BH-white dwarf binary candidate that has been identified in 47\,Tuc \citep{Miller-Jones+2015,Bahramian+2017}. In that case, X-ray and radio data, as well as gravitational microlensing studies, could help to further constrain the BH population in 47\,Tuc \citep{Kains+2018,Zaris+2020}.

\subsection{Applicability of CMC-based results to MUSE data}

In view of the different properties of the observed and simulated binaries, the question arises as to how reliable the detection efficiencies and scaling factors calculated based on the CMC simulations are and whether they are applicable to the real data. In particular, the simulated binaries have on average higher RV amplitudes, which may facilitate their detection and lead to more binaries with high variability probabilities.  In this case, the expected number of binary stars derived from the simulated data would be overestimated, and the actual fraction of binary stars would consequently be underestimated. 

We can roughly estimate the magnitude of this effect by reducing the RV amplitudes in the mock data set. Reducing all amplitudes to 70\% of their value, which corresponds to the ratio of the mean amplitude between the mock and fitted binaries, leads to a 0.2 reduction in the detection efficiency from the previously determined value of $\frac{1}{1.09} = 0.92$ to $0.74$. This would translate to a total binary fraction of 2.9\% instead of 2.4\% in 47\,Tuc. The effect is largely independent of the probability threshold chosen, and although this would represent a substantial increase of the total binary fraction, the uncertainty in the total binary fraction due to an overestimated discovery fraction would still be within the stated uncertainties. 

Another aspect to consider is the applicability of the radial distribution of binary stars in the CMC simulations to extrapolate the binary fraction observed in the MUSE FoV of 47\,Tuc to the total binary fraction in the GC. Since the observed and simulated radial distributions of binary stars agree very well, at least in the central region of 47\,Tuc, where only a minimal increase in the binary fraction towards the center is observed (see Fig.~\ref{fig:binary_fraction_radial_profile}), the extrapolation based on the CMC simulations seems justified. Moreover, there is additional evidence from photometric studies \citep{Milone+2012} that also suggests that the radial dependence of the binary fraction is small.

\section{Summary and conclusions}
\label{sec:conclusion}

Using RVs determined by multi-epoch MUSE spectroscopy, this work examines binary stars in the dense central region of the Galactic GC 47\,Tuc. This cluster is an exceptionally interesting target because it is a particularly old and massive GC, with the latter ensuring good statistics. Estimates of the fraction of binary stars in the cluster have so far been limited to photometric studies of MS binaries, and the distributions of the orbital parameters of binaries in 47\,Tuc and in comparable clusters are largely unexplored. At the same time, specially tailored state-of-the-art simulations of 47\,Tuc provide detailed model predictions of the binary star population, allowing for direct comparison of theory and observation.  

The cleaned MUSE data set consists of 21,699 stars and comprises 245,522 spectra. Each star in the sample has between two and 68 (with a median of 11) reliable RV measurements spanning eight years of observations. After rigorous filtering of stars and the spectra, 708 stars were identified as binary candidates based on their RV variability. This corresponds to a discovery binary fraction in the MUSE FoV of $f_{\mathrm{bin, discovery}} = (3.3\pm1.1)\%$.

The results of mock data tests allow to correct for observational biases and incompleteness. Taking into account the limited sensitivity and lower magnitude limit, the total binary fraction of the cluster is estimated to be $f_{\mathrm{bin, total}} = (2.4\pm1.0)\%$. This measurement is consistent with previous photometric estimates for double MS binaries with high mass ratios \citep{Milone+2012,Ji&Bregman2015}. The relative binary abundances across different stellar types further show a significantly increased binary fraction among blue straggler stars that is approximately three times higher than the cluster average. This result is in line with the theoretical expectation that blue straggler stars form in multistar systems.

For stars with sufficiently many observations, the Keplerian orbital parameters are inferred using the nested sampling package UltraNest \citep{Buchner2021}. 30 binaries with well-constrained orbits, that is, mono- or bimodal posterior period distributions returned by UltraNest, are presented. A cross-match with emission line object and X-ray source catalogs to identify possibly accreting compact binaries reveals four potential X-ray sources and five stars with an excess in H$\alpha$ compared to the fitted model spectra.

The distribution of the inferred binary orbital parameters is interpreted in the context of Cluster Monte Carlo simulations of 47\,Tuc \citep{Ye+2022}. A comparison with these dynamical models reveals a surprising deficit of short-period binaries and binaries with massive companions, which should in principle be the easiest to detect. This result implies a need for further investigation of the initial conditions and orbital evolution of binaries in dense cluster environments and suggests a low number of detectable dark massive companions in binaries with luminous stars in 47\,Tuc.

With the MUSE instrument, a high observational completeness is achieved for stars on the RGB and down to approximately one magnitude below the MS turn-off. However, binaries with white dwarf or lower MS primary stars are below the instrument sensitivity and cannot be detected. In this sense, the search for binaries with MUSE can be seen as complementary to purely photometric studies of lower MS binaries \citep{Milone+2012, Ji&Bregman2015} and near-ultraviolet and X-ray observations of cataclysmic variables \citep[e.g.,][]{Heinke+2005,RiveraSandoval+2018}.

An exciting future avenue is to compare the binary properties of 47\,Tuc with those of other GCs. To date, NGC1850 \citep{Saracino+2023b}, NGC 3201 \citep{Giesers+2019} and 47\,Tuc are among the few GCs observed by MUSE with sufficient epochs to determine the orbital period, eccentricity, and companion mass distributions of binary systems. More observations are needed to extend this analysis to a larger number and variety of GCs. This would allow us to derive trends in the properties of binary stars with cluster properties such as mass, density, age, and metallicity, and would provide valuable information on the dynamic state and evolution of GCs. 

\section*{Data availability}
Binary properties of the systems with well-constrained orbits are listed in Table~\ref{tab:binary_system_properties}. Plots for phase-folded RV curves and MAP orbit models can be found as supplementary material on \href{https://doi.org/10.5281/zenodo.14314046}{Zenodo} for the subset of well-constrained binaries. The RV time series data, binary probabilities, and stellar parameters underlying this study will also be published on \href{https://doi.org/10.5281/zenodo.14314046}{Zenodo}. The MUSE observations analyzed here are available through the \href{https://archive.eso.org/cms.html}{ESO archive}. Other data will be shared on reasonable request to the authors.

\begin{acknowledgements}

The authors thank the anonymous referee for their thorough review and thoughtful and constructive comments. J. Müller-Horn thanks Hans-Walter Rix and Kareem El-Badry for their support of the project and insightful discussions. We thank Daniel Vaz for helpful comments on the manuscript. We acknowledge the support of the Deutsche Forschungsgemeinschaft (DFG) through project DR~281/41-1. C.S.Y. acknowledges support from the Natural Sciences and Engineering Research Council of Canada (NSERC) DIS-2022-568580.
\end{acknowledgements}

%
   \bibliographystyle{aa} 
   \bibliography{references.bib} 
%

\appendix
\section{Appendix}
In table \ref{tab:binary_system_properties}, the binary system properties for the subset of binaries with well-constrained orbits are summarized.

\begin{table*}[ht]
    \centering
    \caption{Properties of the 30 identified binary systems with well-constrained orbits.}
    \label{tab:binary_system_properties}
    \begin{tabular}{c c c c c c c c c l}
    \toprule
        ID & RA & Decl & F606W  & $m_1$  & min. $m_2$  & $P$ &  $K$ & $e$ & Comment \\ 
        & [deg] & [deg]  & [mag] & [M$_\odot$] & [M$_\odot$] & [d] & [km/s] &  & \\ \midrule
        1358 & 5.9975 & -72.1090 & 16.92 & 0.78 & 0.47 & $66.25 ${\tiny$ \pm 0.03$} & $21.4${\raisebox{0.5ex}{\tiny$^{+0.9}_{-0.8}$}} & 0.13{\tiny$ \pm 0.05$} & ~ \\ 
        22875 & 6.0072 & -72.1080 & 18.16 & 0.73 & 0.39 & $5.17 ${\tiny$ \pm 0.01$} & 44.7{\raisebox{0.5ex}{\tiny$^{+3.5}_{-3.3}$}}& 0.04{\raisebox{0.5ex}{\tiny$^{+0.06}_{-0.03}$}} &  \\ 
        26528 & 5.9889 & -72.0966 & 17.94 & 0.75 & 0.16 & $54.52${\raisebox{0.5ex}{\tiny$^{+16.08}_{-8.26}$}} & 9.6{\raisebox{0.5ex}{\tiny$^{+2.0}_{-1.4}$}} & 0.14{\raisebox{0.5ex}{\tiny$^{+0.21}_{-0.10}$}} & ~ \\
        29053 & 5.9744 & -72.1069 & 18.46 & 0.72 & 0.11 & $17.44 ${\tiny$ \pm 0.01$} & 10.3{\raisebox{0.5ex}{\tiny$^{+1.7}_{-1.3}$}} & 0.11{\raisebox{0.5ex}{\tiny$^{+0.16}_{-0.09}$}} & ~ \\
        29629 & 5.9718 & -72.0959 & 17.10 & 0.78 & 0.17 & $7.25 ${\tiny$ \pm 0.01$} & 19.2{\raisebox{0.5ex}{\tiny$^{+1.6}_{-1.5}$}}& 0.08{\raisebox{0.5ex}{\tiny$^{+0.07}_{-0.06}$}} & ~ \\ 
        30718 & 5.9659 & -72.0993 & 18.20 & 0.73 & 0.32 & $41.40 ${\tiny$ \pm 0.02$} & 20.6{\raisebox{0.5ex}{\tiny$^{+1.9}_{-1.4}$}}& 0.37{\tiny$ \pm 0.08$} & ~ \\
        51235 & 6.0438 & -72.0850 & 15.70 & 0.80 & 0.36 & $190.62 ${\raisebox{0.5ex}{\tiny$^{+97.38}_{-3.83}$}}& 13.2{\raisebox{0.5ex}{\tiny$^{+2.0}_{-1.9}$}} & 0.25{\raisebox{0.5ex}{\tiny$^{+0.19}_{-0.17}$}} & ~ \\ 
        53591 & 6.0357 & -72.0931 & 16.97 & 0.78 & 0.47 & $4.90 ${\tiny$ \pm 0.01$} & $50.8${\tiny$ \pm 3.2$}& 0.03{\raisebox{0.5ex}{\tiny$^{+0.05}_{-0.03}$}}& WF4-V04, ELO, X-ray \\ 
        55706 & 6.0294 & -72.0852 & 16.35 & 1.2 & 0.52 & $35.69${\raisebox{0.5ex}{\tiny$^{+0.02}_{-0.06}$}}& $25.3${\tiny$ \pm 3.0$} & 0.42{\raisebox{0.5ex}{\tiny$^{+0.09}_{-0.2}$}} & BSS \\ 
        57067 & 6.0268 & -72.0813 & 16.31 & 1.2 & 0.49 & $2.70 ${\tiny$ \pm 0.01$} & 55.3{\raisebox{0.5ex}{\tiny$^{+5.2}_{-4.2}$}} & 0.28{\raisebox{0.5ex}{\tiny$^{+0.09}_{-0.07}$}} & BSS, ELO\\ 
        61863 & 6.0127 & -72.0840 & 14.21 & 0.82 & 0.15 & $75.20${\raisebox{0.5ex}{\tiny$^{+74.78}_{-0.49}$}} & 11.4{\raisebox{0.5ex}{\tiny$^{+3.4}_{-3.7}$}}& 0.69{\raisebox{0.5ex}{\tiny$^{+0.11}_{-0.15}$}} & bimodal \\ 
        61968 & 6.0114 & -72.0937 & 17.74 & 0.75 & 0.38 & $6.27 ${\tiny$ \pm 0.01$} & $40.0 ${\tiny$ \pm 2.7$} & 0.03{\raisebox{0.5ex}{\tiny$^{+0.05}_{-0.03}$}} & ~ \\
        63549 & 6.0069 & -72.0905 & 16.65 & 1.2 & 0.3 & $15.93${\raisebox{0.5ex}{\tiny$^{+0.67}_{-0.02}$}} & 19.8{\raisebox{0.5ex}{\tiny$^{+3.1}_{-2.6}$}}& 0.11{\raisebox{0.5ex}{\tiny$^{+0.21}_{-0.09}$}} & BSS \\
        65419 & 6.0021 & -72.0868 & 16.94 & 0.77 & 0.72 & $262${\raisebox{0.5ex}{\tiny$^{+2}_{-208.}$}} & 19.2{\raisebox{0.5ex}{\tiny$^{+2.5}_{-2.4}$}}& 0.16{\raisebox{0.5ex}{\tiny$^{+0.22}_{-0.14}$}} & white dwarf cand., bimodal \\
        68514 & 5.9910 & -72.0814 & 16.96 & 0.78 & 0.35 & $4.16 ${\tiny$ \pm 0.01$} & 43.1{\raisebox{0.5ex}{\tiny$^{+4.6}_{-3.7}$}}& 0.15{\raisebox{0.5ex}{\tiny$^{+0.09}_{-0.08}$}} & ~ \\
        70896 & 5.9811 & -72.0835 & 13.53 & 0.82 & 0.35 & $127.36${\raisebox{0.5ex}{\tiny$^{+0.34}_{-0.3}$}} & 14.1{\raisebox{0.5ex}{\tiny$^{+1.0}_{-0.7}$}}& 0.34{\tiny$ \pm 0.06$} & ~ \\
        98813 & 6.0306 & -72.0804 & 15.49 & 0.81 & 0.10 & $33.61 ${\tiny$ \pm 0.03$}& $7.9 ${\tiny$ \pm 0.9$}& 0.12{\raisebox{0.5ex}{\tiny$^{+0.15}_{-0.09}$}} & bimodal \\
        104622 & 6.0138 & -72.0798 & 16.34 & 1.2 & 0.42 & $22.66 ${\tiny$ \pm 0.01$} & $23.4 ${\tiny$ \pm 2.5$} & 0.07{\raisebox{0.5ex}{\tiny$^{+0.1}_{-0.06}$}} & BSS, bimodal  \\ 
        106521 & 6.0083 & -72.0780 & 13.96 & 0.83 & 0.20 & $141.33${\raisebox{0.5ex}{\tiny$^{+0.27}_{-0.34}$}}& 8.1{\raisebox{0.5ex}{\tiny$^{+1.2}_{-1.3}$}} & 0.07{\raisebox{0.5ex}{\tiny$^{+0.13}_{-0.05}$}} & ~ \\ 
        106958 & 6.0070 & -72.0736 & 16.73 & 1.2 & 0.44 & $133.63${\raisebox{0.5ex}{\tiny$^{+209.79}_{-0.98}$}}& 15.7{\raisebox{0.5ex}{\tiny$^{+3.2}_{-2.9}$}}& 0.32{\raisebox{0.5ex}{\tiny$^{+0.17}_{-0.02}$}}& BSS, bimodal \\ 
        107261 & 6.0065 & -72.0734 & 16.21 & 0.79 & 0.33 & $87.60${\raisebox{0.5ex}{\tiny$^{+0.12}_{-20.91}$}}& 14.3{\raisebox{0.5ex}{\tiny$^{+1.8}_{-1.6}$}}& 0.23{\raisebox{0.5ex}{\tiny$^{+0.21}_{-0.13}$}} & ~ \\ 
        110506 & 5.9949 & -72.0799 & 16.76 & 0.77 & 0.36 & $8.94 ${\tiny$ \pm 0.01$} & $33.9 ${\tiny$ \pm 1.6$} & 0.03{\raisebox{0.5ex}{\tiny$^{+0.04}_{-0.02}$}} & SSG, X-ray, ELO  \\  
        112447 & 5.9877 & -72.0738 & 12.49 & 0.80 & 0.27 & $94.26${\raisebox{0.5ex}{\tiny$^{+0.24}_{-0.22}$}}& $11.9 ${\tiny$ \pm 0.6$} & 0.05{\raisebox{0.5ex}{\tiny$^{+0.06}_{-0.04}$}} & ~ \\  
        129292 & 6.0766 & -72.0601 & 14.73 & 0.82 & 0.17 & $27.14${\raisebox{0.5ex}{\tiny$^{+0.06}_{-19.29}$}} & 9.6{\raisebox{0.5ex}{\tiny$^{+1.0}_{-0.9}$}} & 0.07{\raisebox{0.5ex}{\tiny$^{+0.09}_{-0.04}$}} & bimodal \\  
        132746 & 6.0573 & -72.0596 & 17.19 & 0.74 & 0.14 & $8.46${\raisebox{0.5ex}{\tiny$^{+0.01}_{-0.2}$}}& 16.5{\raisebox{0.5ex}{\tiny$^{+1.2}_{-1.5}$}}& 0.06{\raisebox{0.5ex}{\tiny$^{+0.11}_{-0.05}$}} &  SSG, WF2-V32, bimodal\\  
        132812 & 6.0570 & -72.0549 & 18.29 & 0.72 & 0.23 & $52.77${\raisebox{0.5ex}{\tiny$^{+7.94}_{-0.31}$}}& 13.5 {\raisebox{0.5ex}{\tiny$^{+3.7}_{-2.4}$}} & 0.23{\raisebox{0.5ex}{\tiny$^{+0.23}_{-0.16}$}} & ~ \\  
        133593 & 6.0512 & -72.0623 & 17.10 & 0.78 & 0.43 & $16.71 ${\tiny$ \pm 0.01$} & $31.7 ${\tiny$ \pm 1.7$} & 0.06{\raisebox{0.5ex}{\tiny$^{+0.07}_{-0.05}$}} & ~ \\  
        155735 & 6.0718 & -72.0508 & 17.02 & 0.76 & 0.39 & $6.77 ${\tiny$ \pm 0.01$} & 39.9{\raisebox{0.5ex}{\tiny$^{+1.6}_{-1.4}$}}& 0.03{\raisebox{0.5ex}{\tiny$^{+0.04}_{-0.02}$}} & SSG, ELO\\  
        1004589 & 5.9146 & -72.0806 & 17.17 & 0.84 & 0.44 & $35.18${\tiny$ \pm 0.03$}& 25.5{\raisebox{0.5ex}{\tiny$^{+2.0}_{-1.6}$}}& 0.32{\tiny$ \pm 0.04$} & ~ \\
        1009837 & 5.9084 & -72.0830 & 17.84 & 0.88 & 0.49 & $425.44${\raisebox{0.5ex}{\tiny$^{+8.15}_{-7.73}$}}& 11.4{\raisebox{0.5ex}{\tiny$^{+1.3}_{-1.2}$}}& 0.08{\raisebox{0.5ex}{\tiny$^{+0.11}_{-0.06}$}} & ~ \\ \bottomrule
    \end{tabular}
    \tablefoot{The Keplerian parameters were calculated using UltraNest. The columns list the ACS source ID, the right ascension and declination coordinates, the apparent visual magnitude in the F606W filter, the estimated primary and (minimum) companion stellar masses, the period, the RV semi-amplitude, and the eccentricity. Stars showing an excess in H$\alpha$ are marked as emission line objects (ELO) and binaries with bimodal posterior distributions are indicated in the notes column.}
\end{table*}

\subsection{RV uncertainty estimation and calibration}
\label{app:radial_velocities}

RVs were derived by fitting synthetic templates from the library of \cite{Husser+2013} to the observed MUSE spectra. We derived initial uncertainties from the covariance matrix of the Levenberg-Marquardt algorithm used in the spectral fit. As a consistency check, we re-evaluated the RVs by cross-correlating all extracted spectra against synthetic templates (matched based on the best-fit fundamental stellar parameters) from the same library. This generally yielded consistent results within the uncertainties. The treatment of cases where the cross-correlation and spectral fit yielded inconsistent results is described in Appendix~\ref{appendix:quality_cuts}.

We estimated the uncertainties introduced by variations in the wavelength calibration by measuring the velocities of telluric absorption bands in the spectra. The accuracy within a single night was often better than 1\,km/s. To account for these small variations, we conservatively added a 1\,km/s uncertainty to all RV uncertainties in quadrature.

\begin{figure}[t]
    \centering
    \includegraphics[width=\columnwidth]{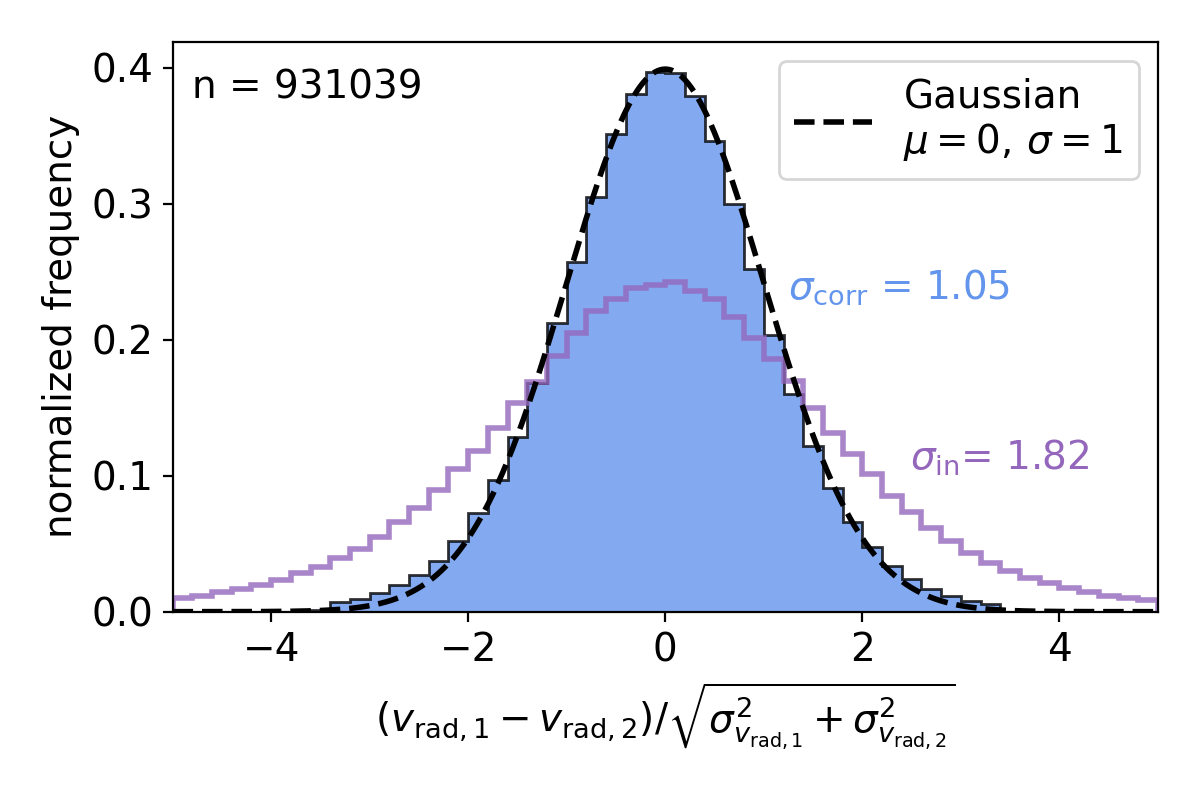}
    \caption{RV differences normalized by their uncertainties (added in quadrature) for stars with multiple measurements. The dashed black line represents the expected standard normal distribution for correctly estimated uncertainties. The observed distribution for 931,039 velocity pairs (purple line) is wider, with a fitted standard deviation of $\sigma_{\mathrm{in}} = 1.82$, indicating underestimated RV uncertainties. After applying S/N-dependent scaling factors to calibrate the uncertainties, the resulting distribution (blue histogram) closely matches the expected normal distribution, with a fitted standard deviation of $\sigma_{\mathrm{corr}} = 1.05$. These calibrated uncertainties were used for the RV variability analysis.}
    \label{fig:rv_uncertainties}
\end{figure}

Examining the distribution of variability probabilities revealed a correlation between the variability probability of a star and its distance to the next brighter star. More specifically, a star was more likely to be classified as a binary star if it had a brighter star in close proximity. There is no direct physical motivation for this trend, as the binary probability should be independent of the distance to neighboring stars. The relation could in principle be explained indirectly in the case of a centrally increasing binary fraction, in which case both the stellar densities and the mean binary probabilities would increase towards the center. However, the correlation persisted even for stars with approximately constant radii from the cluster center. 
The effect is probably due to contamination of the spectra by nearby stars. In this case, the RVs would be determined in part from the contributions of the contaminating stars. This may lead to an increase in RV scatter (on the order of a few km/s) and thus to an increase in binary probability. It appears that the quality criteria for photometric variability and magnitude accuracy are not sensitive enough to remove these stars, so that a low level of contamination may go unnoticed. The increased variability due to contamination is particularly problematic for the analysis of binary star systems, where reliable RVs are crucial. Therefore, we recognize additional RV scatter due to contamination and adjusted the RV uncertainties $\sigma_v$ accordingly for the affected stars. For this purpose, the uncertainties were linearly increased by up to 25\% depending on the distance to the closest star that is 1\,mag brighter than the reference star; $d_{\mathrm{1mag}}$. For all stars with $d_{\mathrm{1mag}} \leq 2''$, the uncertainties were scaled as $\sigma_{v,\mathrm{scaled}} = (1.25 - 0.125 d_{\mathrm{1mag}}) \sigma_v$ and the new uncertainties were used to re-evaluate the variability probabilities. The relation was calibrated to remove the correlation of variability probability with the distance to the next brighter star at a fixed radius from the cluster center. 

Finally, to test the plausibility of the derived uncertainties, we examined stars with multiple observations and compared the differences in velocities $v_{\mathrm{rad,}i}-v_{\mathrm{rad,}j}$ between pairs of measurements with similar S/N values, in the same way as described in \cite{Kamann+2016}. The distribution of normalized velocity differences $$\frac{\left(v_{\mathrm{rad,}i}-v_{\mathrm{rad,}j}\right)} {\sqrt{\sigma_{v_{\mathrm{rad,}i}}^2+\sigma_{v_{\mathrm{rad,}j}}^2}}$$
should follow a Gaussian distribution with a standard deviation of one if the uncertainties are correctly estimated (and neglecting uncertainties of the uncertainties).

We binned these normalized velocity differences by S/N and fit Gaussian distributions to the resulting distributions in each S/N bin. In Fig.~\ref{fig:rv_uncertainties}, we show the distribution of normalized velocity differences for the combined sample from all S/N bins. We found that the normalized velocity differences generally followed normal distributions but with wings slightly wider than $\sigma = 1$, hinting at underestimated uncertainties. To correct for this, we applied an S/N-dependent scaling factor to the uncertainties, ensuring that the normalized velocity differences had a standard deviation of one across all S/N bins. The distribution of velocity differences after scaling the uncertainties is shown in Fig.~\ref{fig:rv_uncertainties} as the solid blue histogram, again for all S/N bins combined.

In principle, this test assumes constant RVs for all stars, meaning that RV variable stars systematically widen the distributions. However, since we found the fraction of binary stars in 47\,Tuc to be very low (see Sect.~\ref{sec:binary_fraction}), we do expect RV variables to have only a minor impact on the widths of the distributions and we also apply a kappa-sigma clipping to remove obvious outliers.


\subsection{Quality cuts}
\label{appendix:quality_cuts}

As summarized in Sect.~\ref{sec:sample_selection}, several quality cuts were applied to the MUSE data set to remove unreliable spectra (and RVs) and avoid spurious variability. The reliability of each spectrum was assessed using several criteria, each flagged as true (1) or false (0). These criteria were then combined in an empirically weighted equation to determine the overall reliability, $R_{\text{total}}$.

First, we applied a S/N threshold; $R_{S/N} = \left(\nicefrac{S}{N} \geq 5\right).$
We also assessed the quality of the cross-correlation using the $r$-statistic as defined by \cite{Tonry_Davis1979} and the full width at half maximum (FWHM), with $R_{\text{cc}} = (r \geq 4 \land \text{FWHM} > 10\,\text{\AA})$.
Next, the uncertainty $\epsilon_{v,\text{cc}}$ of the cross-correlation-derived RV and the uncertainty $\epsilon_{v}$ of the full-spectrum fit RV were required to exceed 0.1 km s$^{-1}$, corresponding to the flags $R_{\epsilon v, \text{cc}}$ and $R_{\epsilon v}$.
We ensured that the measured velocity was within $3\sigma$ of the cluster velocity, $v_{\text{cluster}} = -18\,\mathrm{km\,s}^{-1}$ \citep{Harris1996,Harris2010}, accounting for the cluster velocity dispersion $\sigma_{v,\text{cluster}} = 12.2\,\mathrm{km\,s}^{-1}$ \citep{Baumgardt_Hilker2018} and a typical binary tolerance of 30\,km s$^{-1}$:
$$R_{v} = \frac{|v - v_{\text{cluster}}|}{\sqrt{\epsilon_{v}^{2} + \sigma_{v,\text{cluster}}^{2} + (30\,\text{km}\,\text{s}^{-1})^2}} \leq 3.$$
Finally, we checked if the cross-correlation RV was consistent with the full-spectrum fit RV, within $3\sigma$; with flag $R_{v = v_{\text{cc}}}$.
These criteria were combined into the overall reliability score:
\[
R_{\text{total}} = \frac{(2R_{S/N} + 10R_{\text{cc}} + R_{\epsilon v,\text{cc}} + 3R_{v} + 2R_{\epsilon v} + 5R_{v=v_{\text{cc}}})}{23}.
\]
Spectra with $R_{\text{total}} > 0.8$ were considered reliable.

Furthermore, for each star, non-physical outliers among the RVs, which can occur in cases of low S/N, were identified and excluded. To do this, we iterated through the RV measurements for each star and defined for each velocity the sets $A = {v_i}$ of all RVs and the reduced set $B = A
\setminus {v_x}$ without $v_x$, the RV in question. The velocity $v_x$ with uncertainty $\epsilon_x$ was classified as an outlier if 
    \begin{equation}
        \frac{\sigma_A}{\sigma_B} > 3 \quad \wedge \quad \frac{|v_x-\Tilde{B}|}{\epsilon_{x}+\sigma_B} > 3\,,
    \end{equation}
where $\Tilde{B}$ denotes the median value of the subset $B$ and $\sigma_A, \sigma_B$ are the respective standard deviations of the sets.

Finally, we required that the spectra have (empirically determined) magnitude accuracy of $>90\%$. The magnitude accuracy is a measure of agreement between the recovered broad-band magnitudes from the MUSE spectra and the magnitudes from the photometric source catalog used in the extraction process. A large difference might indicate flux contamination from a nearby star or other issues in the extraction. Following \cite{Kamann+2018}, the magnitude accuracy was calculated as $\left(1 + \nicefrac{\Delta m}{2 \sigma_{\Delta m}} \right)^{-2}$ for each spectrum. $\Delta m$ is the difference between the input and recovered magnitudes and $\sigma_{\Delta m}$ is the standard deviation of the magnitude differences of stars with comparable input magnitudes extracted from the same IFU cube.

\subsection{Variability probability}
\label{appendix:varprob}
To identify binary candidates in the filtered MUSE data set, the variability probability criterion developed by \cite{Giesers+2019} was used, which analyzes the scatter of RVs around the mean value.

For a given star of index $i$, with $m$ measured RVs $v_j$ and corresponding uncertainties $\sigma_j$, the weighted mean $\Bar{v}$ and the $\chi_i^2$ value representing the scatter of the data were computed as
\begin{equation}
    \chi_i^2 = \sum_{j=1}^m \frac{1}{\sigma_j^2} \left(v_j - \Bar{v}\right)^2 \quad \mathrm{and} \quad \Bar{v} = \frac{\sum_{j=1}^m \frac{v_j}{\sigma_j^2}}{\sum_{j=1}^m \frac{1}{\sigma_j^2}}\,. \label{eq:chi-squared}
\end{equation}

For nonvariable stars, the scatter is largely a result of measurement errors, which are assumed to be normally distributed. For a system made entirely of single stars, the observed values of $\chi^2$ should follow a theoretical chi-square distribution with $\nu = m-1$ degrees of freedom (DoF). For stars with significant RV variations that exceed the associated uncertainties, the scatter is dominated by orbital motions. These stars introduce an excess of high $\chi^2$ values compared to the null hypothesis considering only single stars. Therefore, the comparison of the theoretical and observed chi-squared distribution functions allows to identify variable stars.

Based on this premise, \cite{Giesers+2019} define the probability of a star to be variable as
\begin{equation}
    P(\chi_i^2, \nu) = \frac{F(\chi_i^2, \nu_i)_{\mathrm{theoretical}}-F(\chi_i^2, \nu_i)_{\mathrm{observed}}}{1-F(\chi_i^2, \nu_i)_{\mathrm{observed}}}\,.
\end{equation}
Here $F(\chi_i^2, \nu_i)_{\mathrm{theoretical}}$ is the theoretical cumulative distribution function (CDF) for the null-hypothesis that all stars with $\nu_i$ DoF in the sample are single and $F(\chi_i^2, \nu_i)_{\mathrm{observed}}$ is the observed empirical CDF. 
The equation can be generalized to include the data from all DoF by rewriting it as
\begin{equation}
    P(\chi_i^2, \nu_i) = \frac{n\Tilde{F}(\chi_i^2)-k}{n-k}\,, \label{eq:varprob}
\end{equation}

where  $\Tilde{F}(\chi_i^2) = \frac{1}{n} \sum_{\nu} n_{\nu} F(\chi_i^2, \nu)_{\mathrm{theoretical}} $ was introduced to sum up the contributions from different DoFs. $k$ is the number of stars with reduced $\chi^2$ smaller than $\frac{\chi_i^2}{\nu_i}$ and $n$ is the total number stars. 

The minimum required variability probability was set to $P(\chi_i^2) > 0.5$ as a compromise in terms of false-positives. Thus, the bulk of the single stars are removed, while at the same time ensuring that the majority of binaries remain in the sample.

\subsection{Eliminating sources of contamination}
\label{app:contamination}
A source of contamination was identified from the spatial distribution of binary candidates in the MUSE FoV. It was noticed that there is an accumulation of binary candidates in the overlap regions of the MUSE pointings. It turns out that the elevated number of binaries is not a mere consequence of increased sensitivity due to the availability of more epochs in these regions. Instead, for some of the stars there appear to be slight offsets between the RVs from different pointings. More precisely, the stars show no variability when looking at the RVs for each pointing separately, but between the pointings there are approximately constant velocity shifts of several km/s, which result in higher variability probabilities. 
The problem was solved by computing the variability probabilities separately per star and per pointing and then averaging the probabilities per star over all available pointings with at least two measured spectra.\footnote{The NFM pointing and the central pointing were excluded from this analysis since they alone do not span a sufficiently long time to detect variability with periods of weeks to months.} True binary stars should be variable regardless of the pointing in which they were observed, but stars that are variable only because of unphysical offsets between pointings can be removed in this way. There are 607 stars in the data set, corresponding to 2.6\% of all stars and 47\% of the binary candidates, that were no longer classified as variables when the per-pointing variability probability was used instead of the standard approach. Although 2.6\% of stars affected by velocity shifts is a small number in the context of the entire data set, it is extremely important to identify and exclude these stars in order to determine the binary fraction, which is of the same order of magnitude. 

\end{document}